\documentclass[pra,floatfix,amsmath,twocolumn,showpacs,superscriptaddress]{revtex4}
\usepackage[ansinew]{inputenc}
\usepackage{times}
\usepackage{verbatim}
\usepackage{epsfig}
\usepackage{pst-all}
\usepackage{color}
\usepackage{dcolumn}
\usepackage{amsmath}
\usepackage{amsthm}
\usepackage{bm}
\usepackage{layout}
\usepackage{float}
\usepackage{txfonts}
\usepackage{amsfonts}
\usepackage{amssymb}%
\usepackage{subfigure}
\usepackage[ansinew]{inputenc}
\usepackage{amsmath}
\usepackage{amsthm}
\usepackage{float}
\usepackage{amsfonts}
\usepackage{amssymb}

\begin{document}
\title{Global quantum discord and quantum phase transition in XY model}

\author{Si-Yuan Liu }
\affiliation{Institute of Modern Physics, Northwest University, Xian
710069, P. R. China }
\affiliation{Beijing National Laboratory for Condensed Matter Physics,
Institute of Physics, Chinese Academy of Sciences, Beijing 100190, P. R. China}

\author{Yu-Ran Zhang}
\affiliation{Beijing National Laboratory for Condensed Matter Physics,
Institute of Physics, Chinese Academy of Sciences, Beijing 100190, P. R. China}

\author{Wen-Li Yang }
\email{wlyang@nwu.edu.cn }
\affiliation{Institute of Modern Physics, Northwest University, Xian
710069, P. R. China }

\author{Heng Fan}
\email{hfan@iphy.ac.cn}
\affiliation{Beijing National Laboratory for Condensed Matter Physics,
Institute of Physics, Chinese Academy of Sciences, Beijing 100190, P. R. China}
\affiliation{Collaborative Innovation Center of Quantum Matter, Beijing, P. R. China}

\date{\today}

\begin{abstract}
We study the relationship between the behavior of  global quantum correlations
and quantum phase transitions in XY model. We find that the two
kinds of phase transitions in the studied model can be characterized by the
features of global quantum discord (GQD) and the corresponding quantum correlations.
We demonstrate that the maximum of the sum of all the nearest neighbor bipartite GQDs
is effective and accurate for signaling the Ising quantum phase transition,
in contrast, the sudden change of GQD is very suitable
for characterizing another phase transition in the XY model. 
This may shed lights on the study of properties of quantum correlations in 
different quantum phases. 
\end{abstract}


\maketitle

\section{introduction}
The recent development in quantum information theory \cite{key-1}
has provided much insight into quantum phase transitions \cite{key-2}.
In particular, using the quantum correlations to investigate quantum
phase transitions has drawn much attention and has been successful
in characterizing a number of critical phenomena of great interest.
For example, entanglements measured by concurrence, negativity, geometric
entanglement, von Neumann entropy, mutual information and quantum
discord are studied in several critical systems \cite{key-3,key-4,key-5,key-6,key-7,key-8,key-9}.
From the previous literature, we know that the concurrence shows a
maximum at the critical points of the transverse field Ising model
and XY model \cite{key-4}, and the von Neumann entropy diverges
logarithmically at the critical point \cite{key-3}. The quantum
critical phenomena in the XY model can also be characterized by the
divergence of the concurrence derivative or the negativity derivative
with respect to the external field parameter \cite{key-5,key-6}.
Furthermore, recent studies show that entanglement spectra can be used to
describing quantum phase transitions \cite{key-10,key-11}. 
The structure of the correlations is shown to be related with the
quantum critical phenomena \cite{key-15,key-16}.
On the other hand, fidelity and the fidelity susceptibility of the
ground state can also be used as a good tool for detecting numerous
phase transition points in some critical systems \cite{key-12,key-13,key-14}.
Notably, the methods from quantum information may also play a key
role for the topology of many-body system and the phase transition
of only one spin \cite{fan1,fan2,fan3}.    

Additionally, the quantum discord can be used
to describing the quantum phase transitions in some critical systems \cite{key-17,key-18,key-19,key-20}.
It is worth noting that some of the investigations performed so far
have indicated that quantum discord is more sensible than entanglement
in revealing quantum critical points \cite{key-21}, even for systems
that are not at zero temperature \cite{key-22}. As we all know,
there are several multipartite promotions of quantum discord \cite{key-23,key-24,key-25,key-26,key-27,key-28}.
The global quantum discord (GQD) proposed by Rulli and Sarandy is
an widely accepted one \cite{key-28}, which can be seen as a generalization
of symmetric bipartite quantum discord. There is an interesting question
that if the GQD can be used as a good tool to characterizing the quantum
phase transitions in the typical critical systems. On the contrary,
there is another interesting question that if our understanding of
the quantum phase transitions can tell us some useful information
of the behavior of GQD in the critical systems. Fortunately, the answers
of these questions are yes.

In this paper, we investigate the relationship between the behavior
of global quantum discord and quantum phase transitions in the XY
model \cite{key-29,key-30}. Considering the local convertibility,
the phase diagram of XY model can be divided into three phases, which we label phase
1A, phase 1B and phase 2 \cite{key-29}. The are two kinds of quantum
phase transitions in XY model, the phase transition between phase
1A and phase 1B and the second order phase transition between phase
1 and phase 2. In order to provide a good description of these quantum
phase transitions, we analysis the behavior of total global quantum
discord (GQD), the sum of all the bipartite GQDs and the residual
GQD \cite{key-31}. We will show that the sum of all the nearest
neighbor bipartite GQDs is effective and accurate for signaling the
Ising quantum phase transitions between phase 1 and phase 2. Moreover,
it is worth noting that the sudden change of GQD is very suitable
for characterizing the phase transitions between phase 1A and phase 1B.
On the other hand, since GQD can be seen as a kind of physical
resource for quantum information processing, the nature of these
quantum phase transitions can tell us some useful features of it,
such as the maximum points and sudden changes.

The paper is organized as follows: In Sec. II, we introduce the XY
model and the definitions of GQD, the sum of all the nearest
neighbor bipartite GQDs
and the residual GQD. In Sec. III, we study the behavior of these
quantum correlations and provide some figures of the three kinds of
correlations and the corresponding derivative, it shows that these
correlations can be used to characterizing the quantum phase transitions
effectively and accurately. At the same time, using the nature of
quantum phase transitions, we can get some useful features of the
behavior of global quantum discord. In Sec. IV, we give some conclusions
and discussions.

\section{The brief introduction of the XY model and global quantum discord \label{II}}

\begin{figure}[t]
\includegraphics[width=0.32\textwidth]{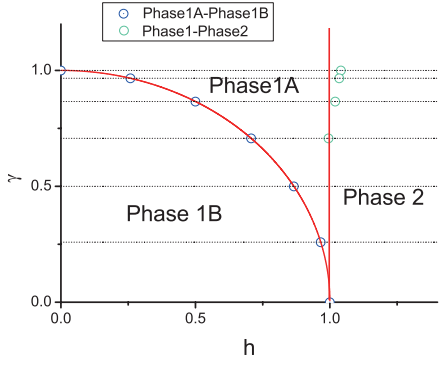}
\caption{Phase diagram of the XY model. Three phases are labelled as phase 1A, phase 1B and phase 2.
For different values of $h$ in our method, the blue circles represent the critical points for phase transition
between phase 1B and phase 1A, and green circles represent the critical points for phase transition
between phase 1A and phase 2.
}
\end{figure}

We study quantum phase transitions in one dimensional XY-model \cite{key-32} by
using the method from quantum information theory. The Hamiltonian
for our model is as follows:
\begin{eqnarray}
H=-\sum_{i=0}^{N-1}\left\{ \frac{J}{2}[\left(1+\gamma\right)\hat{\sigma}_{i}^{x}\hat{\sigma}_{i+1}^{x}+\left(1-\gamma\right)\hat{\sigma}_{i}^{y}\hat{\sigma}_{i+1}^{y}]+h\hat{\sigma}_{i}^{z}\right\},
\label{1}
\end{eqnarray}
with $N$ being the number of spins in the chain, $\hat{\sigma}_{i}^{m}$
the $i$th spin Pauli operator in the direction $m=x,y,z$ and periodic
boundary conditions assumed. The XX model and transverse field Ising model
thus correspond to the the special cases for this general class of
models. For the case that $\gamma\rightarrow0$, our model reduces
to XX model. When $\gamma=1$, the model reduces to transverse
field Ising model. For simplify, here we take $J=1$ and
the parameter $h$ is associated with the external transverse magnetic
field. For $h=1$, a second-order quantum phase transition takes place
for any $0\leq\gamma\leq1$. In fact, there exists additional structure
of interest in phase space beyond the breaking of phase flip symmetry
at $h=1$. It's worth noting that there exists a circle, $h^{2}+\gamma^{2}=1$,
on which the ground state is fully separable. According to the previous
literature, this circle can be seen as a boundary between two differing
phases which are characterized by the presence and absence of parallel
entanglement \cite{key-33,key-34,key-35,key-36}. In fact, for each fixed $\gamma$,
the system is only locally non-convertible when $h^{2}+\gamma^{2}>1$.
Now we can divide the system into three separate phases, phase 1A,
phase 1B and phase 2, where the ferromagnetic region is now divided
into two phases defined by their differential local convertibility.
These results are summarized in ``phase-diagram'' Fig.1.
 Consideration of differential local
convertibility separates the XY model in three phases, which we label
phase 1A, phase 1B and phase 2. The phase transition from phase 1
to phase 2 is the second-order phase transition. The green points
around the critical line are the critical points $h_{c}$ obtained
by our method as $N\rightarrow\infty$. 
Phase 1 has two distinct regions A and B, the boundary is a quarter of a circle,
$h^{2}+\gamma^{2}=1$. The phase transition from phase 1A to phase
1B can be seen as a first-order phase transition when we consider
the GQD as a kind of order parameter. The blue points on the boundary
are the critical points obtained by our method. 
It is obvious that our method is very accuracy for characterizing
the phase transitions.

For understanding the relationship between the global quantum correlations
and quantum phase transitions in XY-model, we study the behavior
of global quantum discord carefully. The global quantum discord is a
measure of multipartite quantum correlation, which can be seen as
a symmetric generalization of bipartite quantum discord to multipartite
cases. As a well-defined multipartite quantum correlation, the GQD
is always non-negative and symmetric with respect to subsystem exchange.
Considering its applications, GQD has been shown to be useful
in many areas, such as quantum communication and quantum phase transitions \cite{key-28,key-37,key-38,key-39,key-40}.
In detail, GQD can play a role in quantum communication, in the sense
that its absence means that the quantum state simply describes a classical
probability multi distribution. That is to say, it allows for local
broadcasting of correlations \cite{key-38}. On the other hand,
the global discord has been proved to be useful in the characterization
of quantum phase transitions. In previous literature, the behavior
of GQD in some typical critical systems, such as the Ashkin-Teller
spin chain, transverse field Ising model, open-boundary XX model
and cluster-Ising model have been studied \cite{key-39,key-40}.
It is worth noting that both of the transverse field Ising model
and XX model can be seen as a special case of XY model, the
nature of quantum phase transitions is more complicated in this case.
So there are two interesting questions: $\left(1\right)$ How the behavior
of GQD is like in this model; $\left(2\right)$ If the global discord
can be used to describing the complicated and richness critical phenomenon
in this model. Fortunately, the answers are yes.

\begin{figure*}[t]
 \centering
\subfigure{\includegraphics[width=0.195\textwidth]{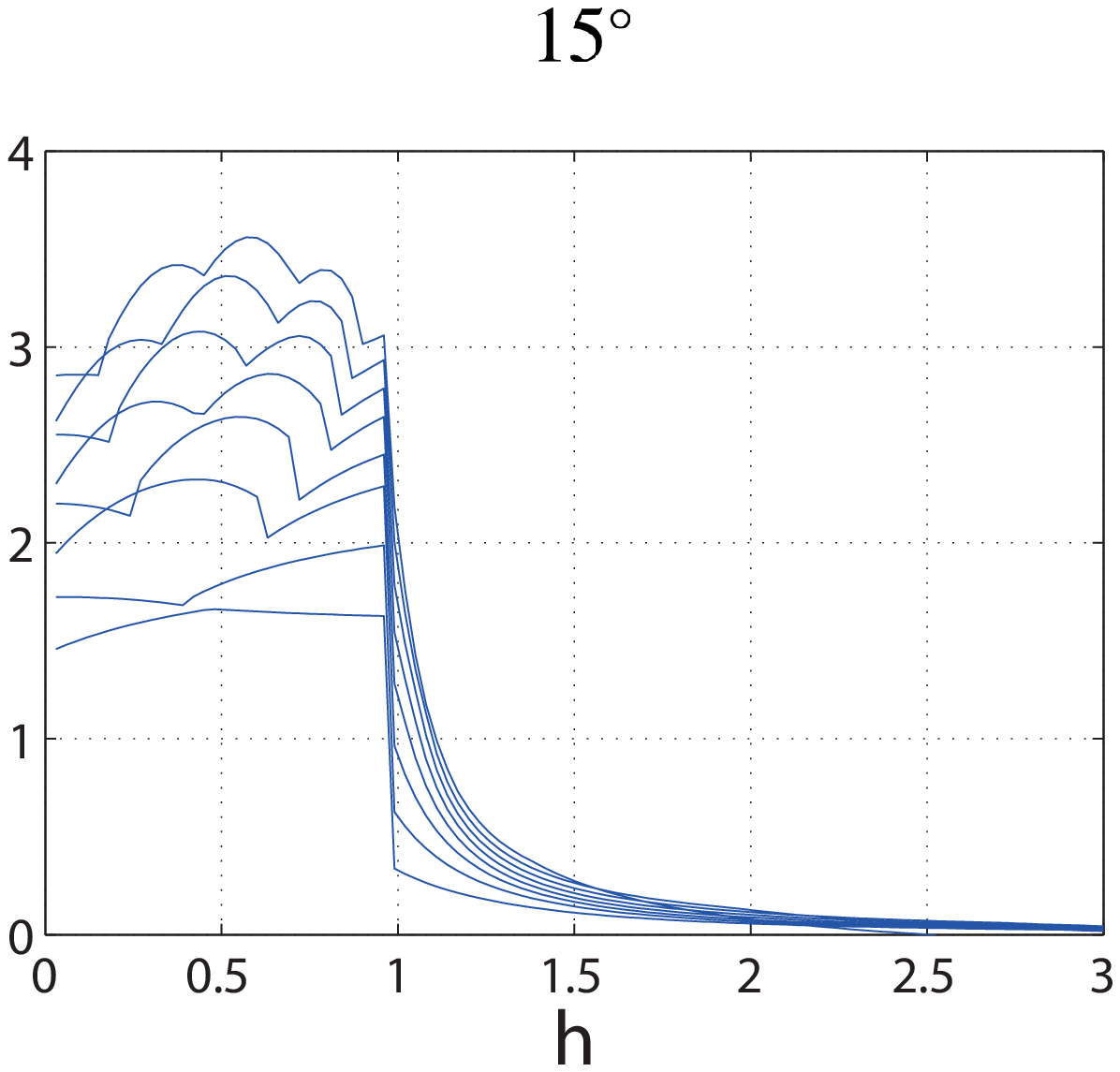}}
\subfigure{\includegraphics[width=0.195\textwidth]{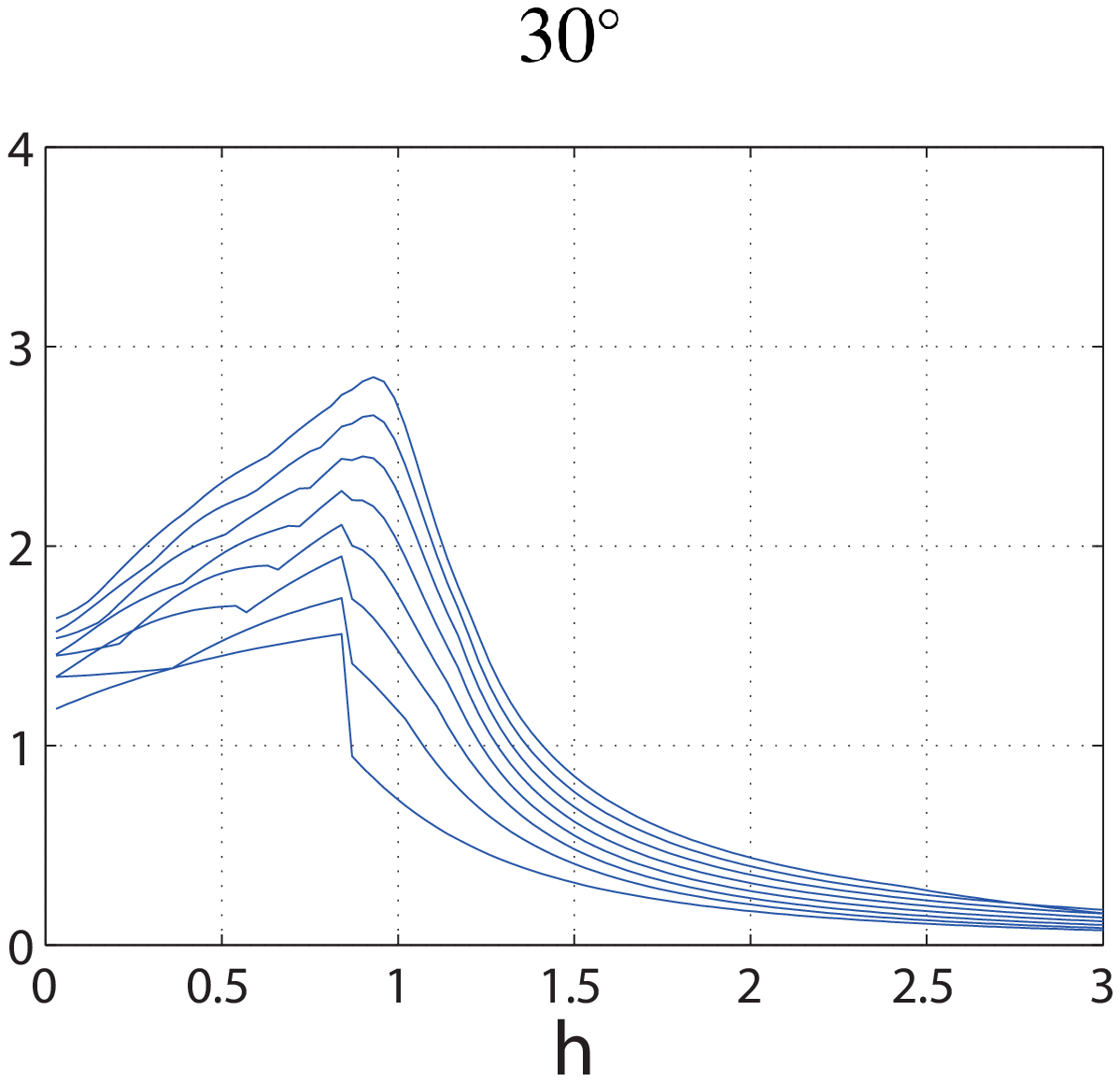}}
\subfigure{\includegraphics[width=0.195\textwidth]{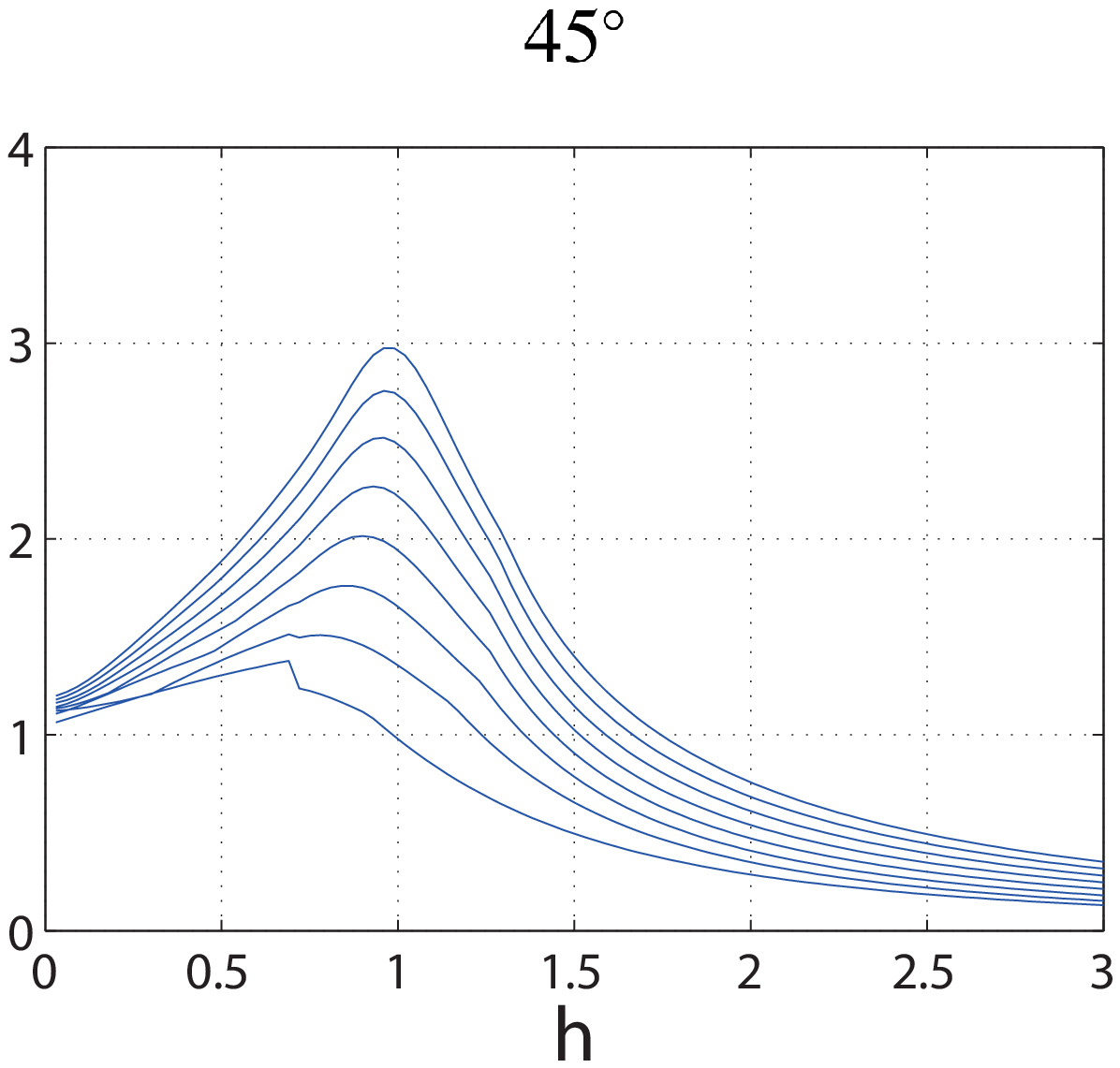}}
\subfigure{\includegraphics[width=0.195\textwidth]{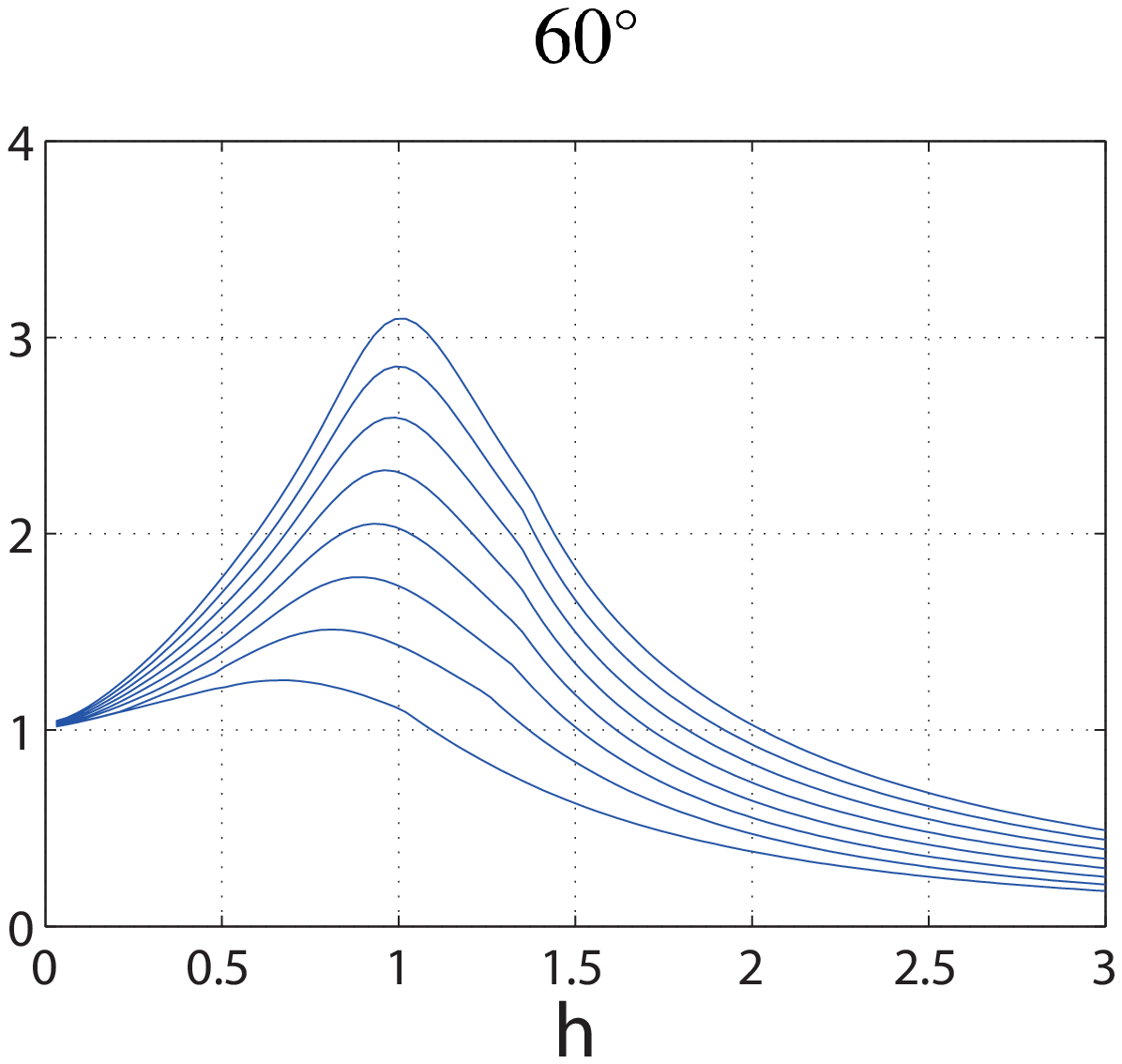}}
\subfigure{\includegraphics[width=0.195\textwidth]{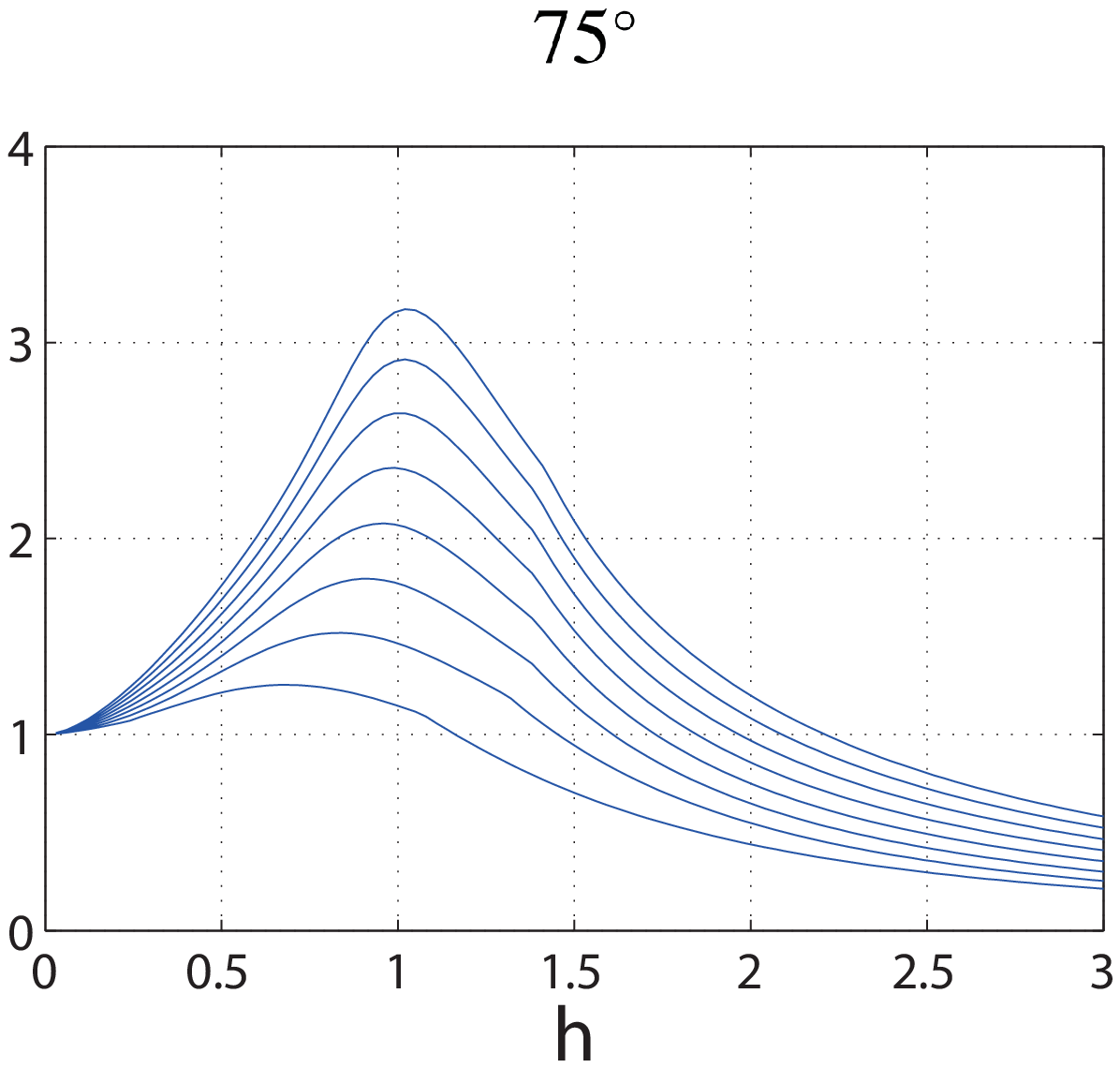}}
\subfigure{\includegraphics[width=0.195\textwidth]{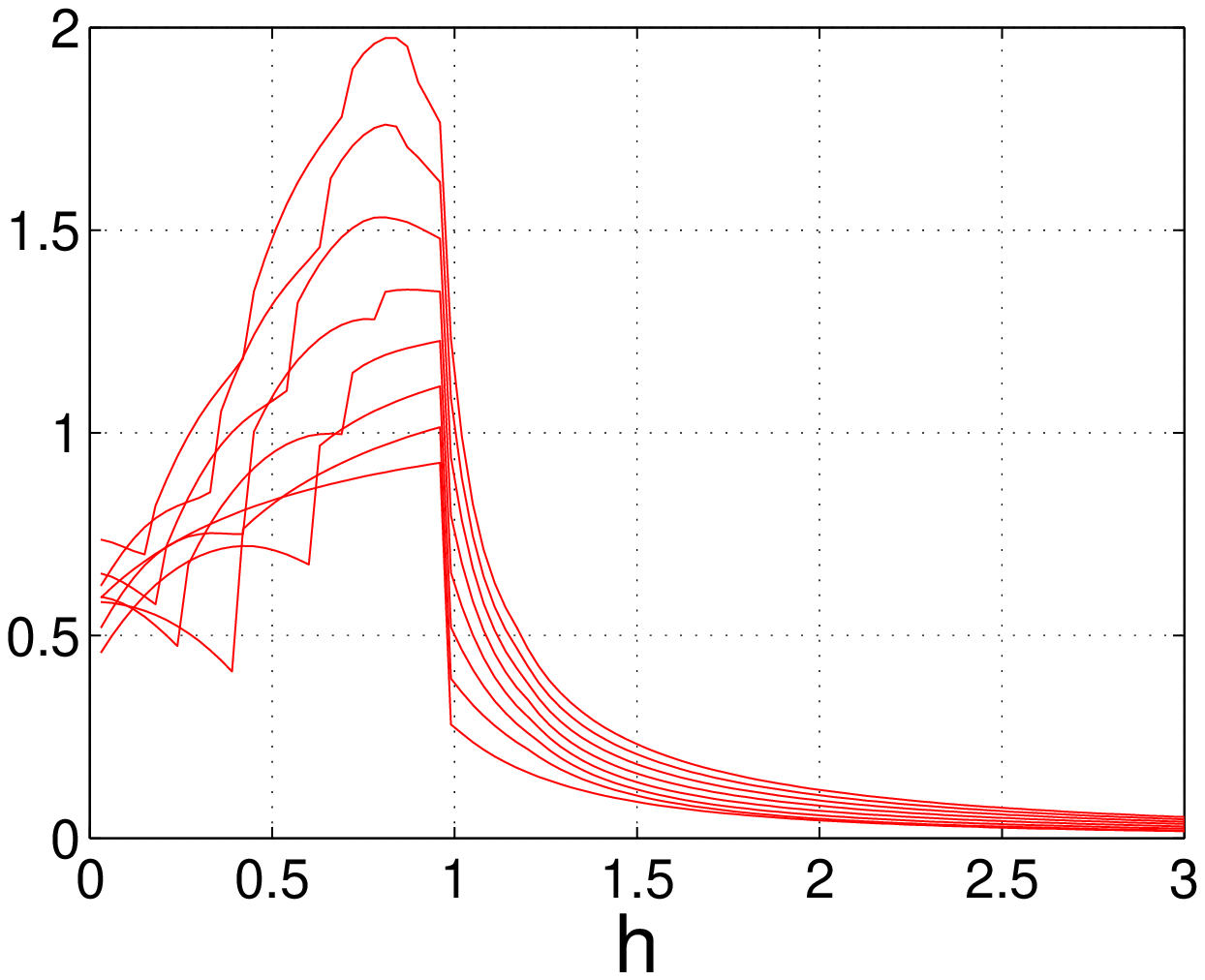}}
\subfigure{\includegraphics[width=0.195\textwidth]{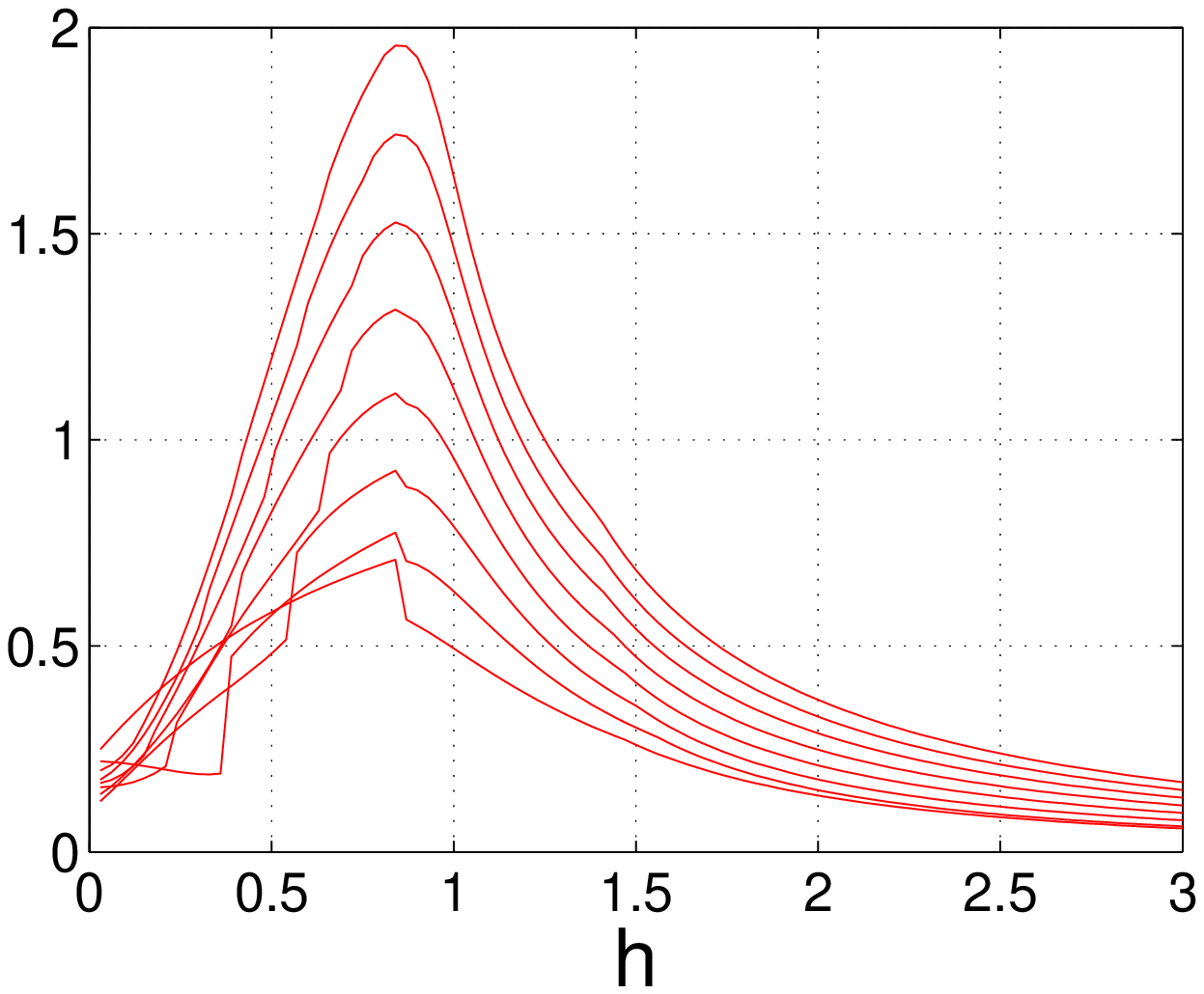}}
\subfigure{\includegraphics[width=0.195\textwidth]{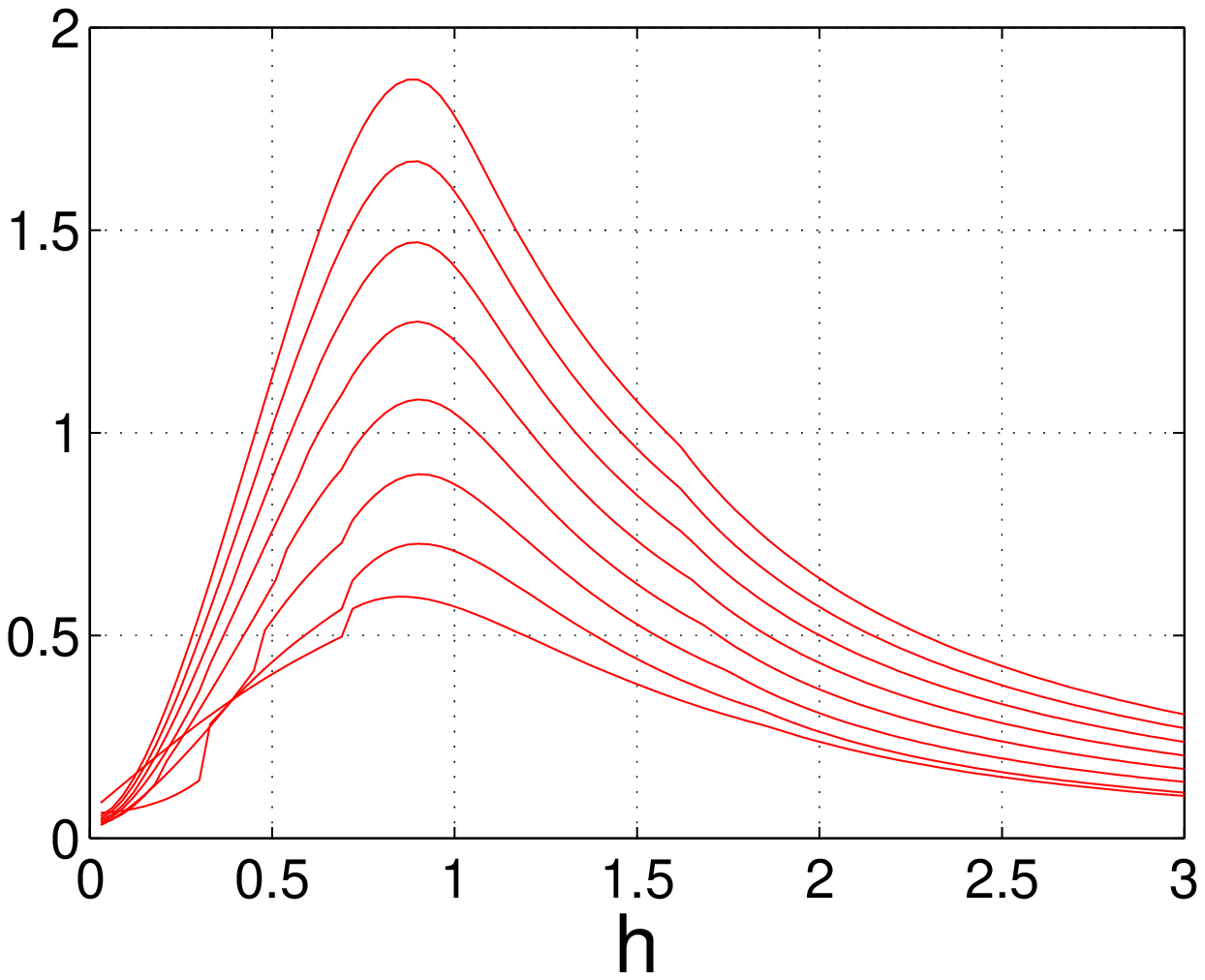}}
\subfigure{\includegraphics[width=0.195\textwidth]{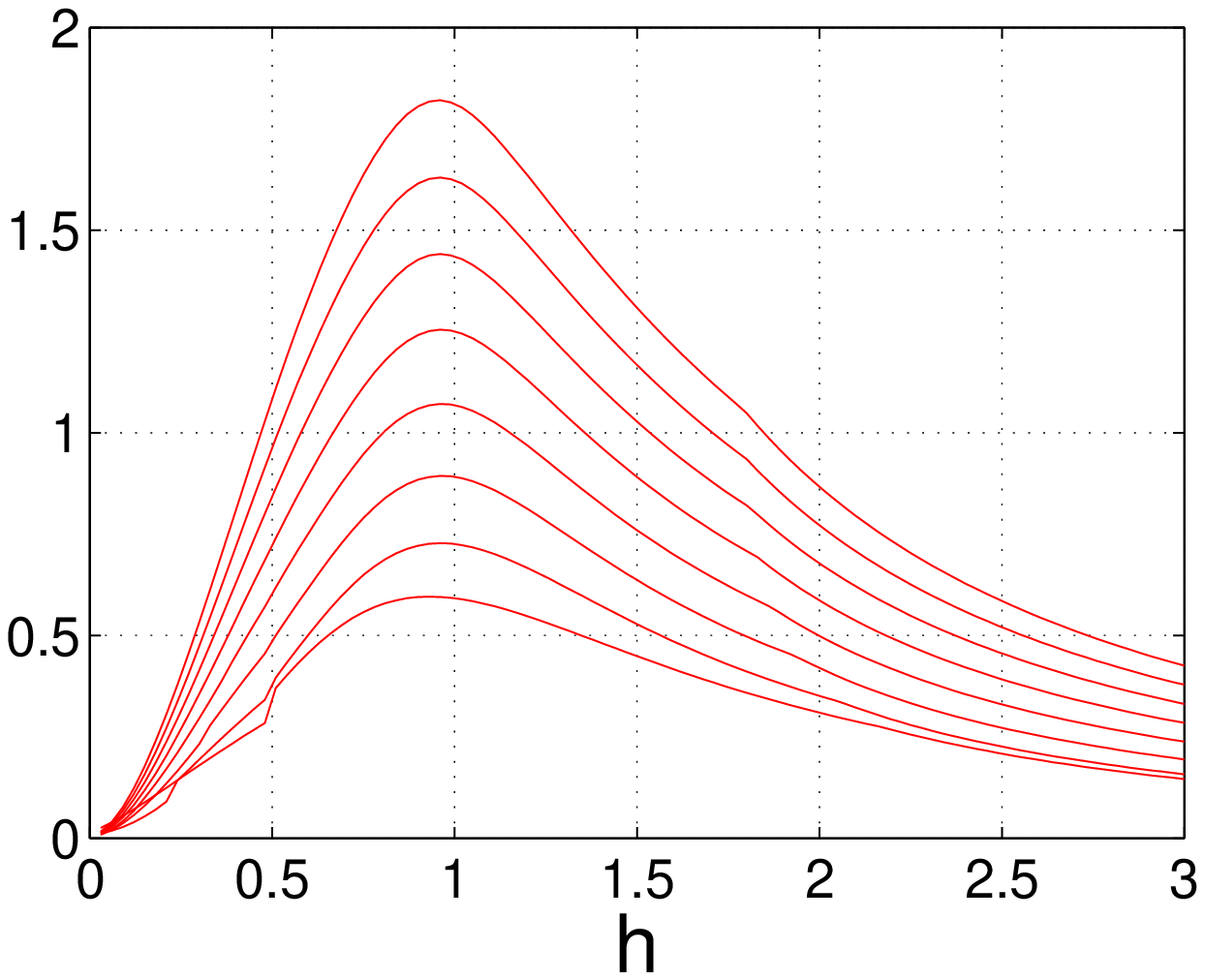}}
\subfigure{\includegraphics[width=0.195\textwidth]{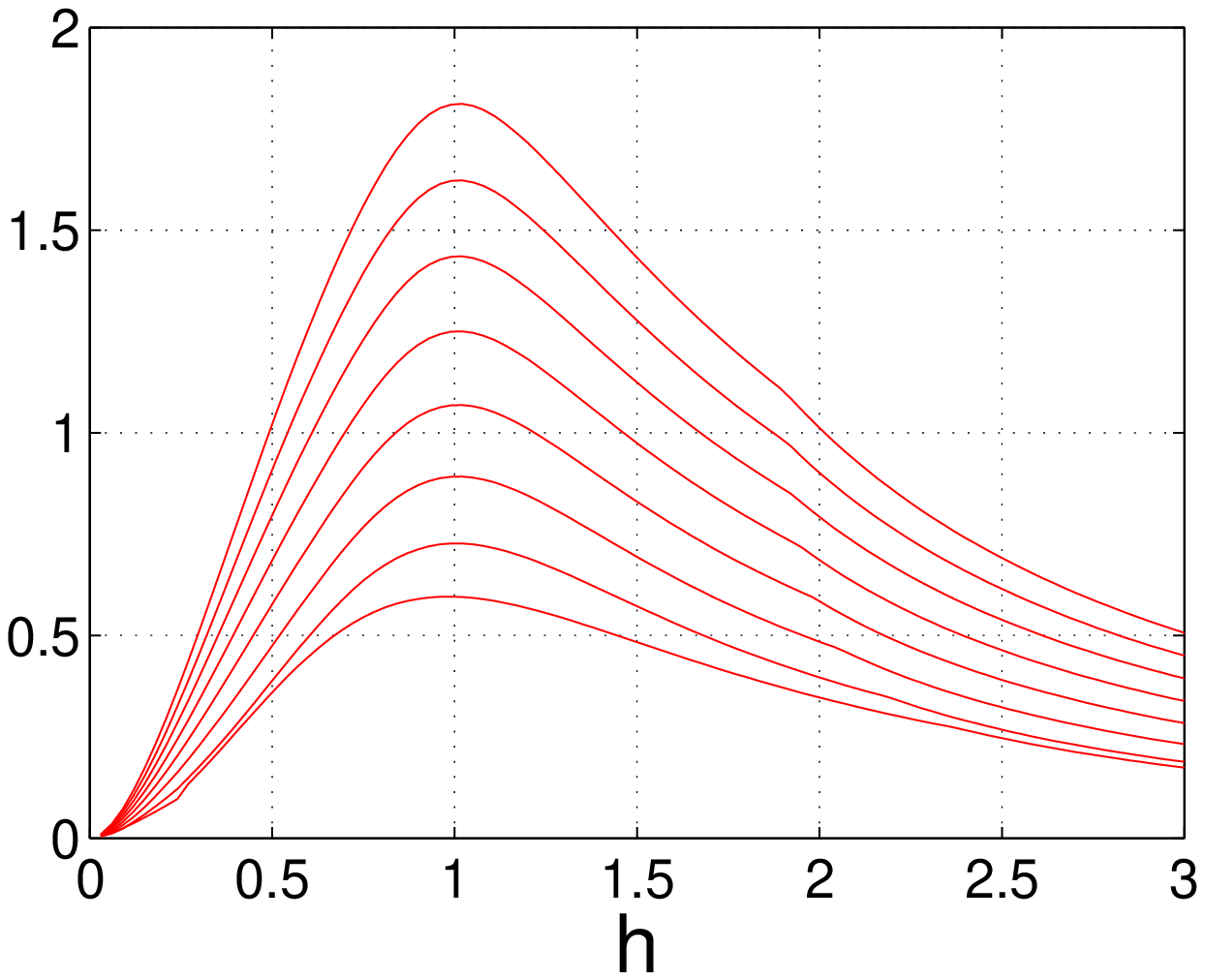}}
\subfigure{\includegraphics[width=0.195\textwidth]{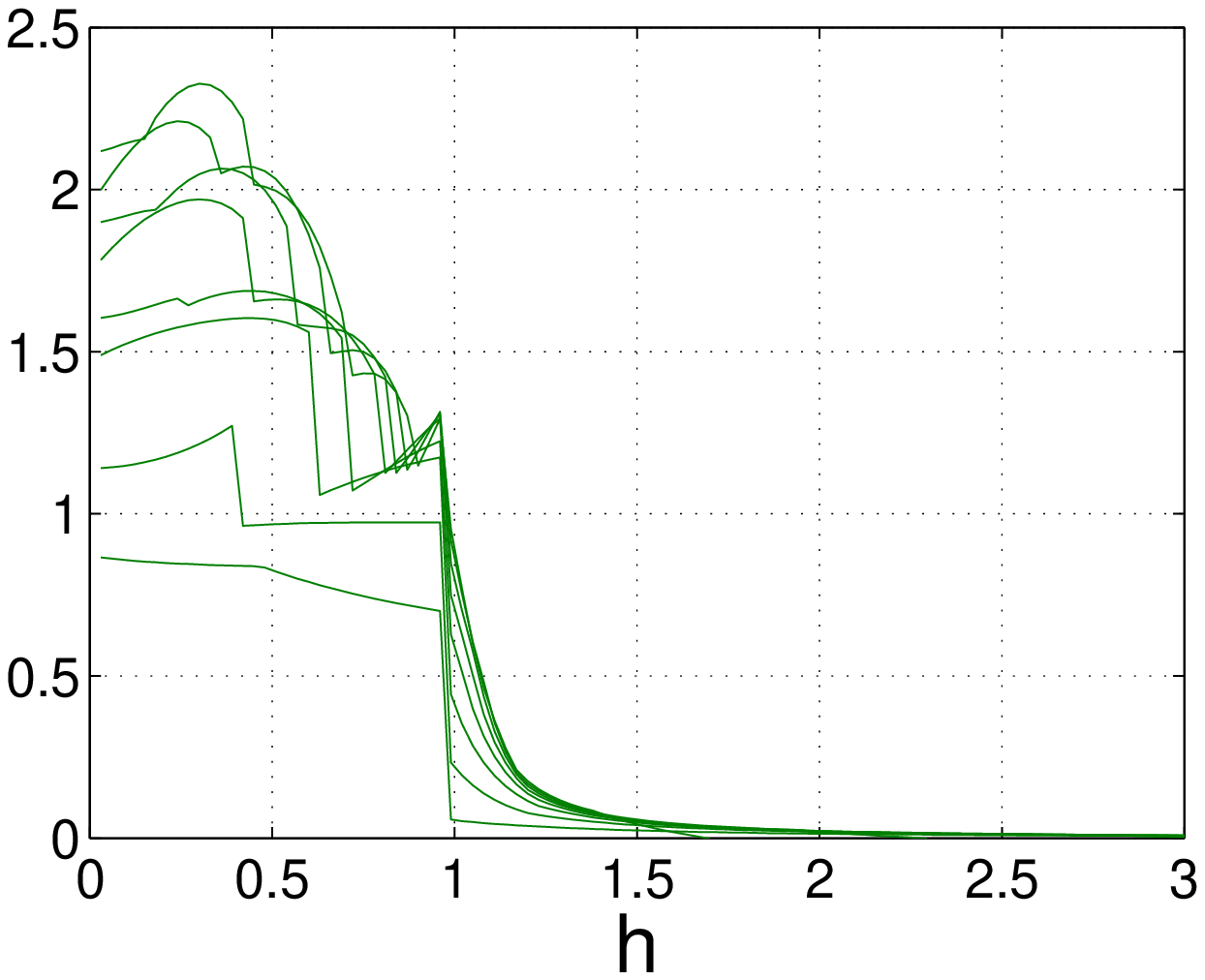}}
\subfigure{\includegraphics[width=0.195\textwidth]{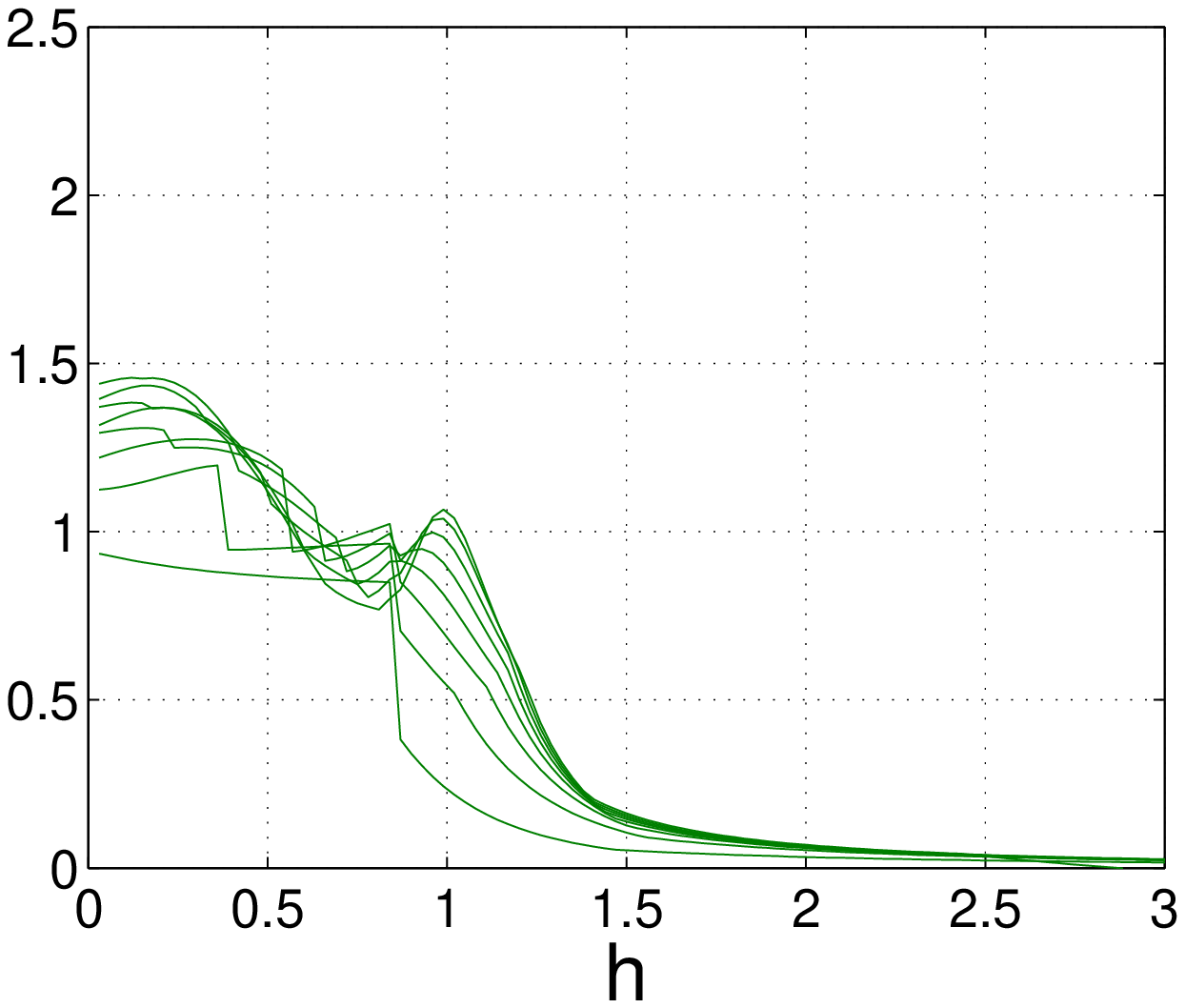}}
\subfigure{\includegraphics[width=0.195\textwidth]{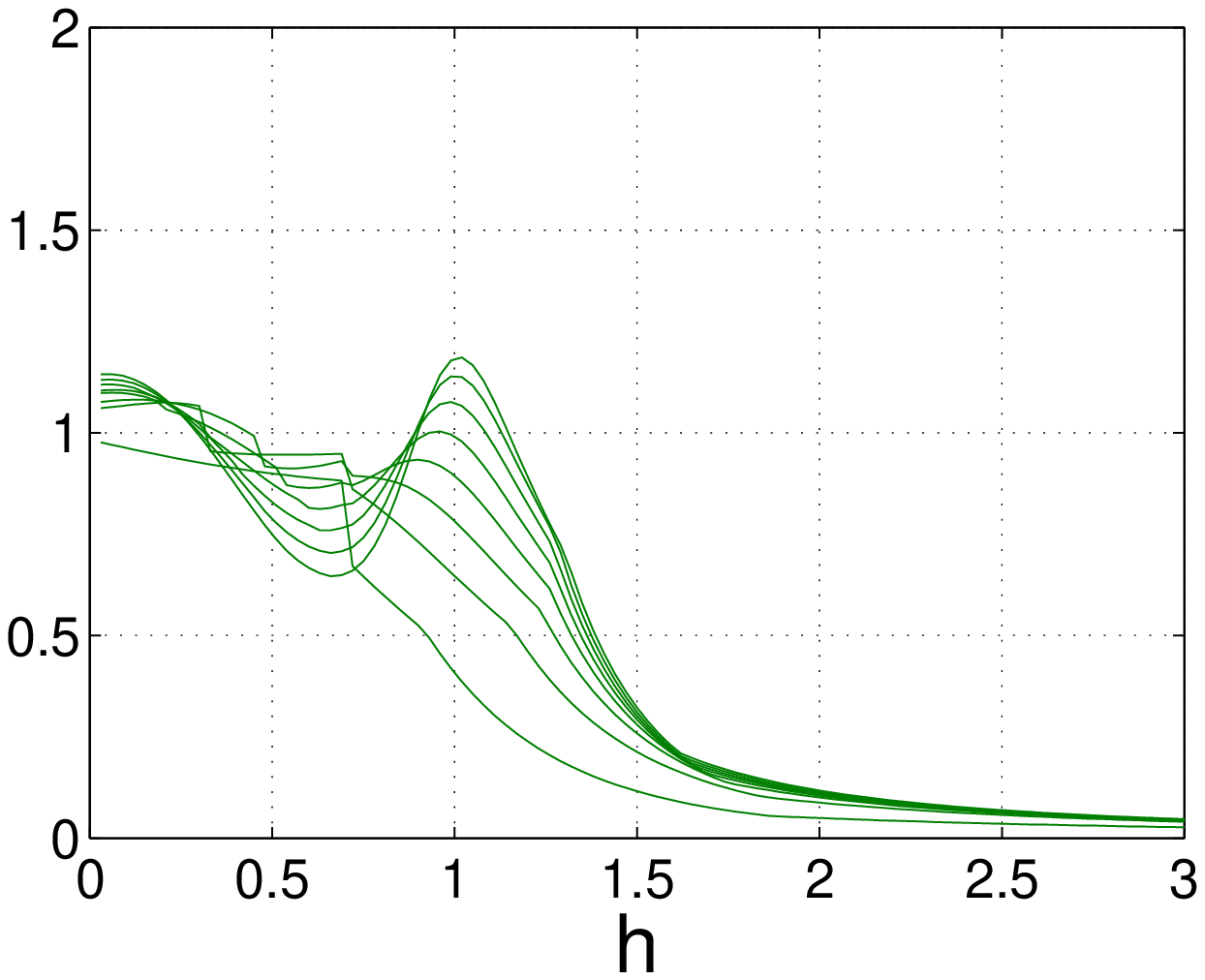}}
\subfigure{\includegraphics[width=0.195\textwidth]{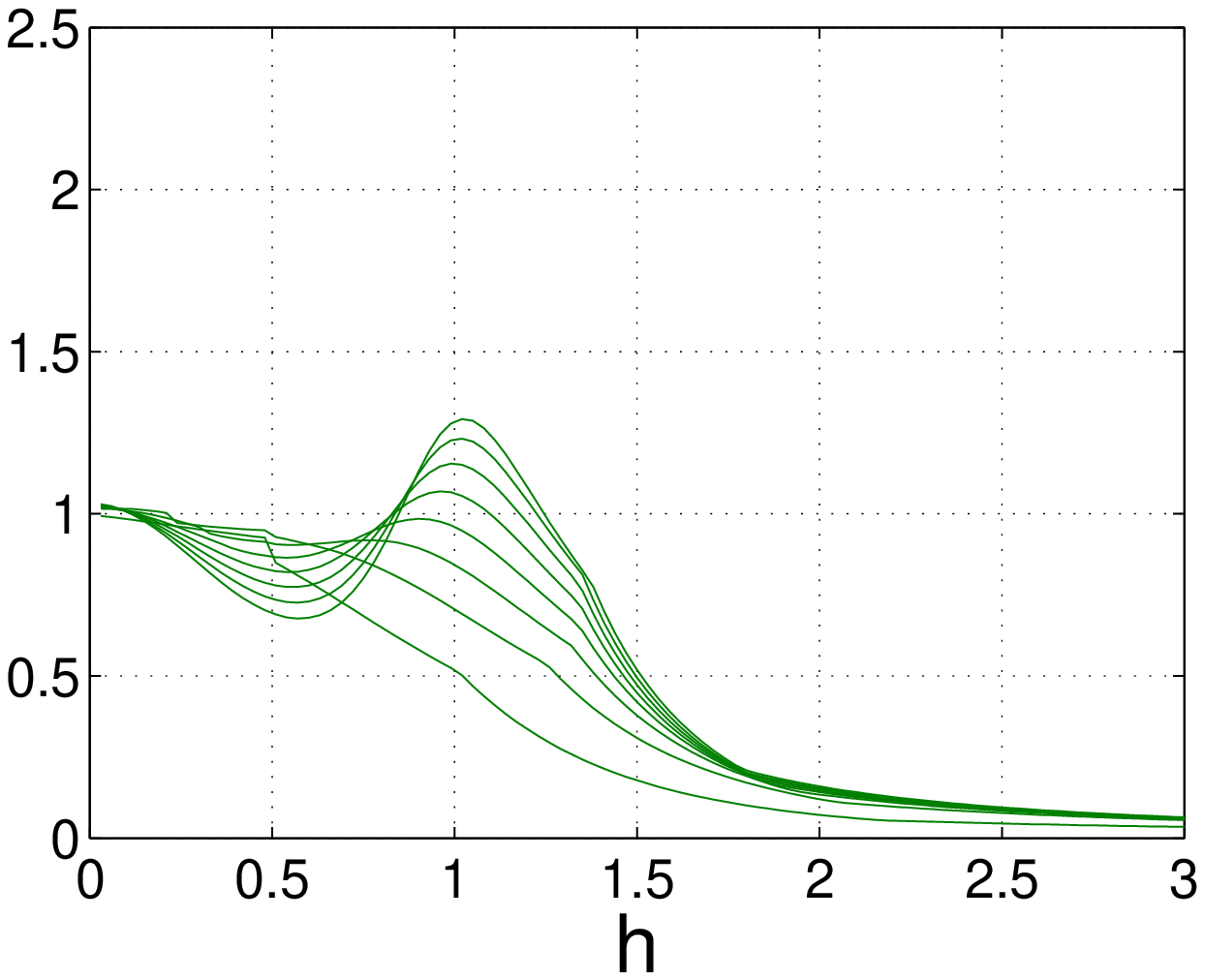}}
\subfigure{\includegraphics[width=0.195\textwidth]{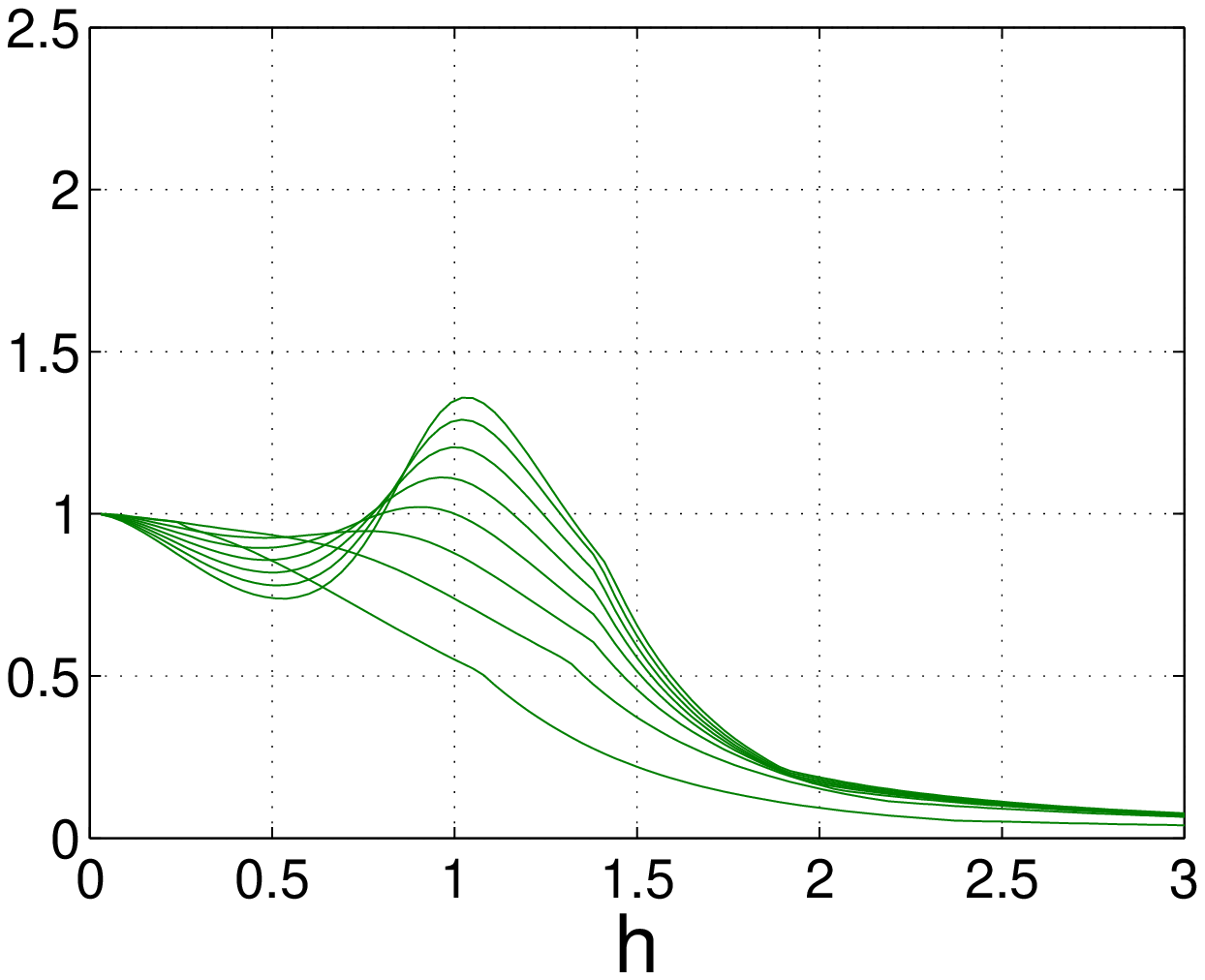}}
\caption{Total GQD, the sum of all nearest neighbor bipartite
GQDs and the residual GQD against $h$, when $\gamma=\sin\theta$,
($\theta=15^{\circ},30^{\circ},45^{\circ},60^{\circ},75^{\circ}$). The figures
in each column correspond to a specific angle: from first column
to last column, the $\theta$ increases from $15^{\circ}$ to $75^{\circ}$.
The first line is about GQD, the second line shows the
sum of all nearest neighbor bipartite GQDs, the last
line is corresponding to the residual GQD.
In each figure
we study rings with $L = 3,\cdots,10$, from bottom to top.}
\end{figure*}

In order to study the behavior of global discord in our model, we
first review the definition of it. The definition of global quantum
discord is a generalization of bipartite symmetric quantum discord.
Consider a $N$-partite system A$_{1}$, A$_{2}$, ... , A$_{N}$
(each of them is of finite dimension), the GQD of state $\rho_{A_{1}A_{2\cdots}A_{N}}$
is defined as follows:
\begin{eqnarray}
D\left(A_{1}:\cdots:A_{N}\right)\equiv\underset{\Phi}{\min}\left[I\left(\rho_{A_{1}\cdots A_{N}}\right)-I\left(\Phi\left(\rho_{A_{1}\cdots A_{N}}\right)\right)\right],
\label{2}
\end{eqnarray}
where
$\Phi\left(\rho_{A_{1}\cdots A_{N}}\right)={\sum}_{k}\Pi_{k}\rho_{A_{1}\cdots A_{N}}\Pi_{k}$,
with \{$\Pi_{k}=\Pi_{A_{1}}^{j_{1}}\otimes\cdots\otimes\Pi_{A_{N}}^{j_{N}}$\}
representing a set of local measurements and $k$ denoting the index
string $\left(j_{1}\cdots j_{N}\right)$.
In Eq.~(\ref{2}), the multipartite mutual information $I\left(\rho_{A_{1}\cdots A_{N}}\right)$
and $I\left(\Phi\left(\rho_{A_{1}\cdots A_{N}}\right)\right)$ are
given by
\begin{eqnarray}
I\left(\rho_{A_{1}\cdots A_{N}}\right)=\sum_{k=1}^{N}S\left(\rho_{A_{k}}\right)-S\left(\rho_{A_{1}\cdots A_{N}}\right),
\\
I\left(\Phi\left(\rho_{A_{1}\cdots A_{N}}\right)\right)=\sum_{k=1}^{N}S\left(\Phi\left(\rho_{A_{k}}\right)\right)-S\left(\Phi\left(\rho_{A_{1}\cdots A_{N}}\right)\right),
\label{4}
\end{eqnarray}
where
$\Phi\left(\rho_{A_{k}}\right)={\sum}_{k^{'}}\Pi_{A_{k}}^{k^{'}}\rho_{A_{k}}\Pi_{A_{k}}^{k^{'}}$.
Based on the definition of GQD, we can define two corresponding multipartite
correlations \cite{key-31}. The sum of GQDs between two nearest
neighbor particles is defined as
\begin{eqnarray}
D\left(A_{1}:A_{2}\right)+D\left(A_{2}:A_{3}\right)+\cdots+D\left(A_{L-1}:A_{L}\right),
\label{6}
\end{eqnarray}
$L$ is the size of our system. This correlation represents all nearest neighbor bipartite GQDs contained in the critical system. Then,
similar to the definition of tangle as a measure of residual multipartite
entanglement, we can define the residual GQD corresponding to the
second monogamy relation,
\begin{eqnarray}
D_{R}^{L}\equiv D\left(A_{1}:\cdots:A_{L}\right)-\sum_{i=1}^{L-1}D\left(A_{i}:A_{i+1}\right).
\label{7}
\end{eqnarray}
It is a measure for residual multipartite quantum correlation, namely,
contributions to quantum correlation beyond pairwise GQD. This measure
of residual multipartite quantum correlation describes the total
quantum correlation except for all nearest neighbor interaction
of quantum correlations. In most cases, since the bipartite GQDs do
not increase under the discard of subsystems, the second monogamy
relation holds. That is to say, the residual GQD is greater than or
equal to zero. In other words, the system contains non-zero long-range
correlation.

In order to calculate the global quantum correlations mentioned above,
we first reformulate GQD as \cite{key-40}:
\begin{eqnarray}
D(A_{1}:\cdots:A_{L})&=&\underset{\{ \hat{\Pi^{k}}\} }{\min}\left\{ \sum_{j=1}^{L}\sum_{l=0}^{1}{\tilde{\rho}_{j}^{ll}}\log_{2}\tilde{\rho}_{j}^{ll}-\sum_{k=0}^{2^{L}-1}\tilde{\rho}_{T}^{kk}\log_{2}\tilde{\rho}_{T}^{kk}\right\} \nonumber\\
&+&\sum_{j=1}^{L}S(\rho_{j})-S(\rho_{T})\label{8}
\end{eqnarray}
with $\tilde{\rho}_{T}^{kk}=\langle\bm{k}|\hat{{R}}^{\dag}\rho_{T}\hat{{R}}|\bm{k}\rangle$ and
$\tilde{\rho}_{j}^{ll}=\langle l|\hat{{R}}_{j}^{\dag}\rho_{j}\hat{{R}}_{j}| l\rangle$, where
$\hat{\Pi}^{k}=\hat{R}|\bm{ k}\rangle\langle \bm{k}|\hat{R}^{\dag}$ are the multi-qubit
projective operators. Here $\left\{ |\bm{k}\rangle\right\}$ are separable eigenstates of
$\bigotimes_{j=1}^{L}\hat{\sigma}_{j}^{z}$, 
and $\hat{R}$ is a local $L$-qubit rotation: $\hat{R}=\bigotimes_{j=1}^{L}\hat{R}_{j}(\theta_{j},\phi_{j})$
with $\hat{R_{j}}(\theta_{j},\phi_{j})=\cos\theta_{j}\hat{I}+i\sin\theta_{j}\cos\phi_{j}\hat{\sigma}_{y}+i\sin\theta_{j}\sin\phi_{j}\hat{\sigma}_{x}$
acting on the $j$-th qubit.

This formula greatly reduces the computational efforts needed to evaluate
GQD.

\section{The behavior of global quantum discord and quantum phase transitions
in the XY model\label{III}}
\begin{figure}[t]
\subfigure[]{\includegraphics[width=0.225\textwidth]{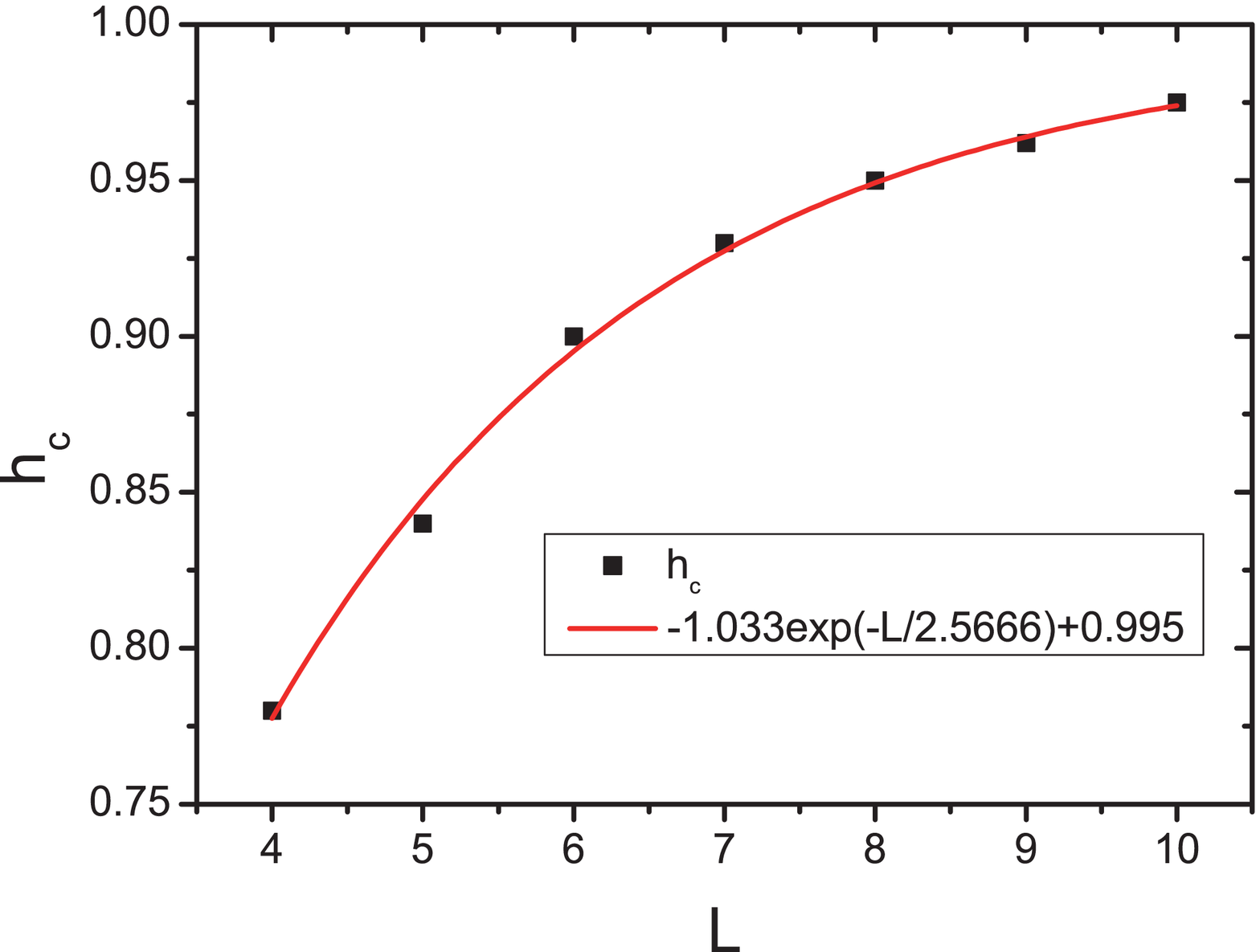}}
\subfigure[]{\includegraphics[width=0.225\textwidth]{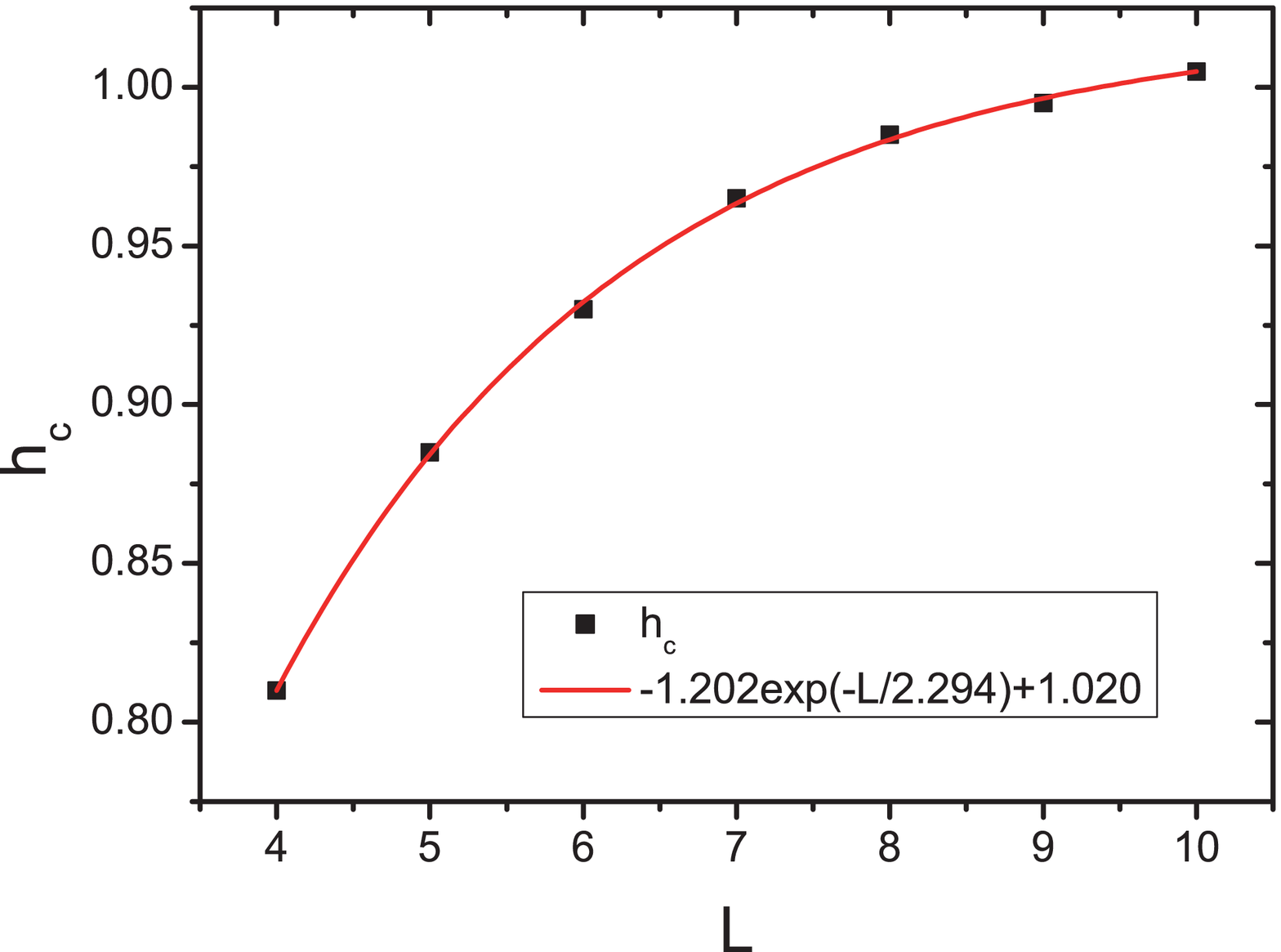}}
\caption{Finite-size scaling  analysis of the  critical points for $\theta=45^{\circ}$ and $\theta=60^{\circ}$ in the
phase transition between phase 1A and phase 2.
(a) is for $\theta=45^{\circ}$ with accuracy 0.00922 and (b) is for $\theta=60^{\circ}$ with accuracy 0.00263.}
\end{figure}

In this section, we analysis the behavior of the three kinds of multipartite
quantum correlations mentioned above in XY model. In order to see
the features of these correlations more clearly, we provide some figures
of them and their derivatives. It shows that the property of these
quantum correlations can be applied to characterizing the two kinds
of quantum phase transitions effectively and accurately. From another
perspective, since GQD can be considered as a kind of physical
resource in quantum information processing, our knowledge about
quantum phase transitions can also help us to understand and predict
the behavior of these correlations better. That is to say, the study
of quantum critical systems can help us make a better understanding
of the quantum correlations and quantum information processing.

Now we consider the nature of these quantum correlations as $h$
changes at the zero temperature similar as \cite{key-31}. 
Fig.2 shows the total GQD, the sum of all nearest neighbor bipartite
GQDs $D\left(A_{1}:A_{2}\right)+D\left(A_{2}:A_{3}\right)+\cdots+D\left(A_{L-1}:A_{L}\right)$
and the residual GQD $D\left(A_{1}:\cdots:A_{L}\right)-\sum_{i=1}^{L-1}D\left(A_{i}:A_{i+1}\right)$
as a function of $h$, when $\gamma=\sin\theta$,
($\theta=15^{\circ},30^{\circ},45^{\circ},60^{\circ},75^{\circ}$).
The figures in each column correspond to a specific angle; from first column to last column,
the $\theta$ increases. The figures in first line are about GQD, the figures in second line show
the sum of all nearest neighbor bipartite GQDs, the figures in last line are corresponding to
the residual GQD. When $\theta=90^{\circ}$, our model reduces to the transverse
field Ising model. We study rings with $L=3,\cdot\cdot\cdot,10$,
from bottom to top.

First of all, we consider the case that $\gamma=\sin75^{\circ}=0.966$ as shown in the
last column of Fig.2.
It shows that the sum of all nearest neighbor bipartite GQDs reaches
a maximum at nearly the critical point $h=1$, which is more suited
to be used to describing the second-order quantum phase transition than
the total GQD. If we consider GQD as a resource for quantum information
processing, these figures tell us that we can get the most resource
for quantum information and computation tasks when the second-order
quantum phase transition occurs. That is to say, in order to obtain
more physical resource for quantum information tasks from this system,
we just need to adjust the external transverse magnetic field parameter
$h$ at the critical point. On the other hand, the phase 1 ($h\leq1$) can be divided into
two phases, phase 1A and phase 1B. The boundary is a circle $h^{2}+\gamma^{2}=1$,
on which the ground state is fully separable. From these figures,
we find an interesting fact that the GQD sudden changes in some
points. In fact, the sudden change which occurs at $h=\cos75^{\circ}=0.259$
can characterize the phase transition between phase 1A and phase
1B very accurately, since this point is just on the boundary $h^{2}+\gamma^{2}=1$.
From another point of view, when we observed a phase transition like
this, there must be a sudden change occurs. In other words, we can
get a sudden change of these global quantum correlations just by adjusting
corresponding parameter to a appropriate value. If we consider GQD as a kind of order parameter, the phase transition can be seen
as a first-order phase transition. Now we know that the GQD can
be used to detecting both the first-order and second-order phase transitions
in our model.

The fourth column of Fig.2 shows the case that $\gamma=\sin60^{\circ}=0.866$.
Similar as the case that $\theta=75^{\circ}$, the sudden change
which occurs at $h=\cos60^{\circ}=0.5$ can characterize the first-order
phase transition very perfectly. The $\sum_{i=1}^{L-1}D\left(A_{i}:A_{i+1}\right)$
still reaches a maximum at nearly critical point $h=1$, which
detecting the second-order phase transition accurately.

From these figures, we can find that there are many sudden changes
of GQD, only the rightmost sudden change characterizes the phase transition
between phase 1A and phase 1B. Other sudden changes reflect the
level-crossings that redefine the ground state of the system (which
are evident from the spectrum of the model). It is obvious that the
sudden change will be more apparent when we consider the smaller angle
$\theta$ or smaller system size $L$. That is to say, we can use
this method to detect the quantum phase transitions easier
and more effective in these cases. As $\theta$ decreases, the two kinds
of critical points become closer.

In particular, when $\theta=0^{\circ}$, our model reduces to XX model.
In this case, the two kinds of phase transitions both occur at $h=1$.
For low magnetic field the GQD displays a step-wise behavior, jumps
occurring in correspondence of the level-crossings that redefine
the ground state of the system (which are evident from the spectrum
of the model). That is, GQD tracks the structural changes in the ground
state of the spin system as $h$ varies. Comparing all these figures,
we find that as $\theta$ decreases, the GQD curves become more
complicated. When we consider the case that $\theta$ is large enough,
the maximum of the sum of all nearest neighbor bipartite GQDs
can characterize the second-order phase transition accurately. On
the other hand, the phase transition between phase 1A and phase 1B
can be described by the rightmost sudden change of GQD. When $\theta$
is small enough, both phase transitions can be characterized by the
rightmost sudden change of GQD. The sudden change will be more apparent
when we consider the smaller angle $\theta$ or smaller system size
$L$.

To get rid of the finite-size effect in the phase transitions between phase 1A
and phase 2, we exemplify the scaling analysis
with the cases that $\theta=45^{\circ}$ and $\theta=60^{\circ}$ in
Fig.3. From Fig.3 $\left(a\right)$, we find that when $\theta=45^{\circ}$,
the critical point labeled $h_{c}$ obtained by our method tends to
0.995 in the thermodynamic limit. The data can be fit to $h_{c}=-1.033\times\mbox{\ensuremath{\exp}}\left(-L/2.5666\right)+0.955$
with accuracy 0.00922. From Fig.3 $\left(b\right)$, we find that when $\theta=60^{\circ}$,
the critical point labeled $h_{c}$ tends to 1.020 as $N\rightarrow\infty$.
The data can be fitted exponentially as $h_{c}=-1.202\times\exp(-L/2.294)+1.020$
with accuracy 0.00263. For the large system, the critical point obtained
by our method is very accurate.

\begin{figure}[t]
\subfigure{\includegraphics[width=0.225\textwidth]{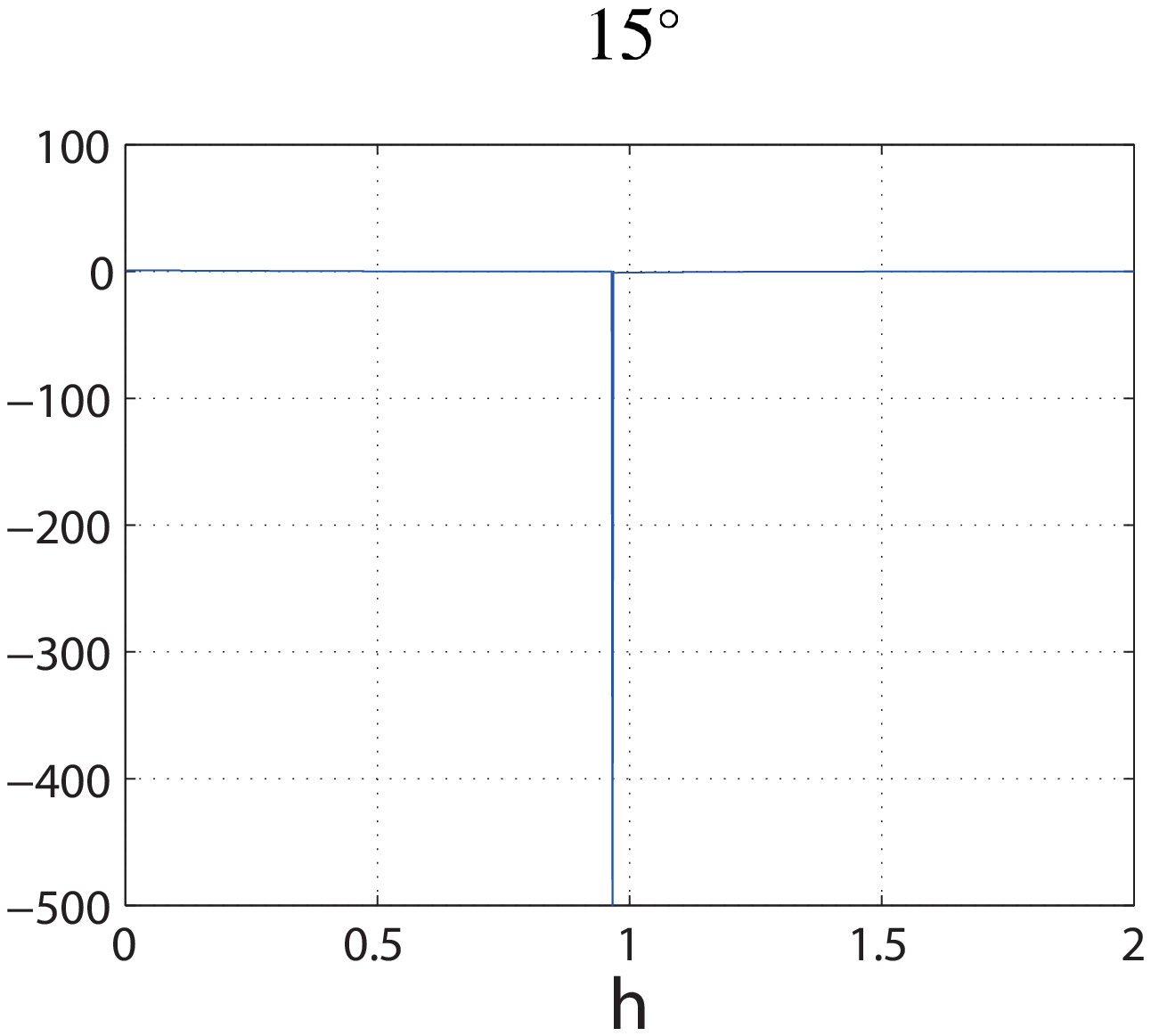}}
\subfigure{\includegraphics[width=0.225\textwidth]{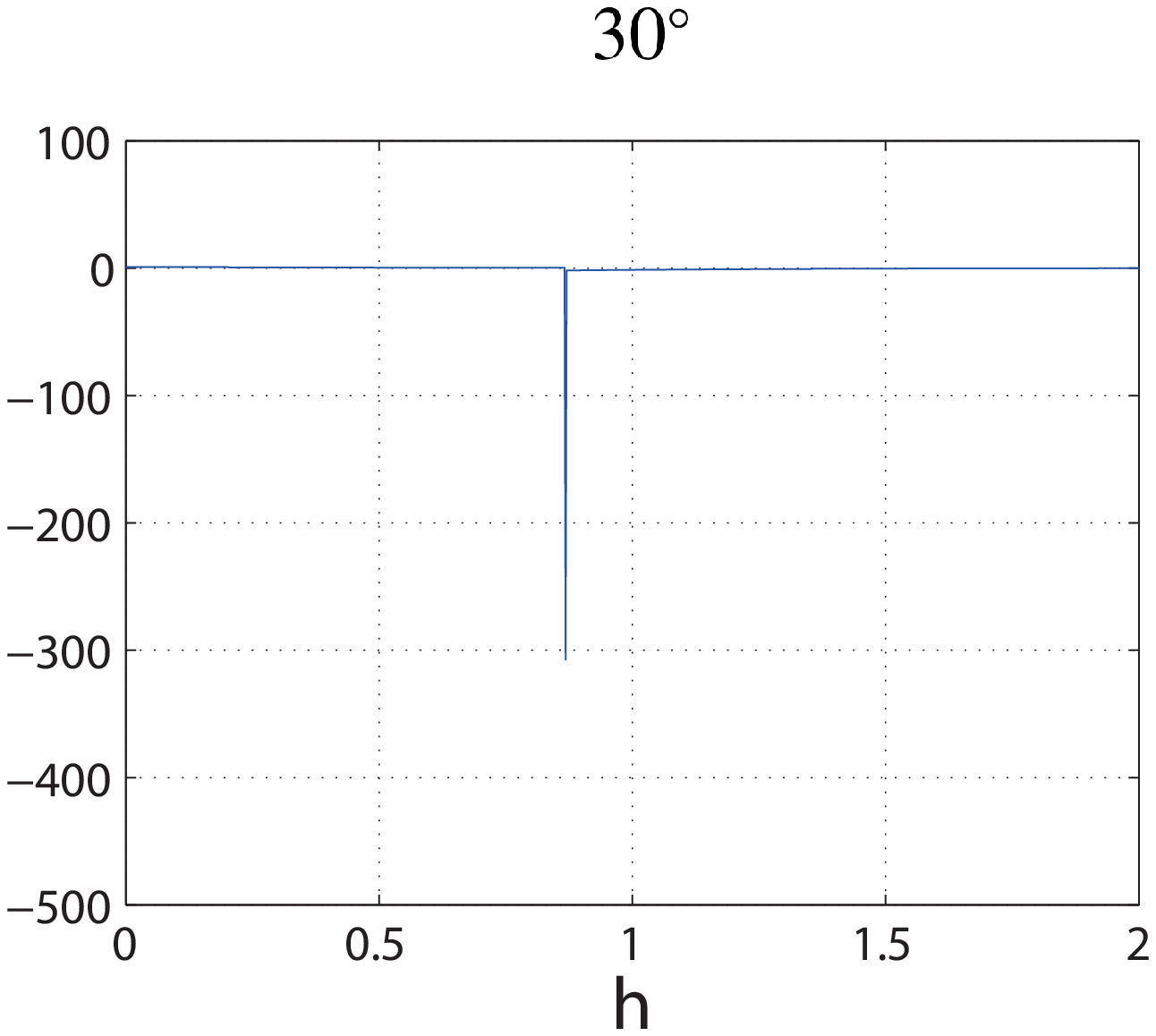}}
\subfigure{\includegraphics[width=0.225\textwidth]{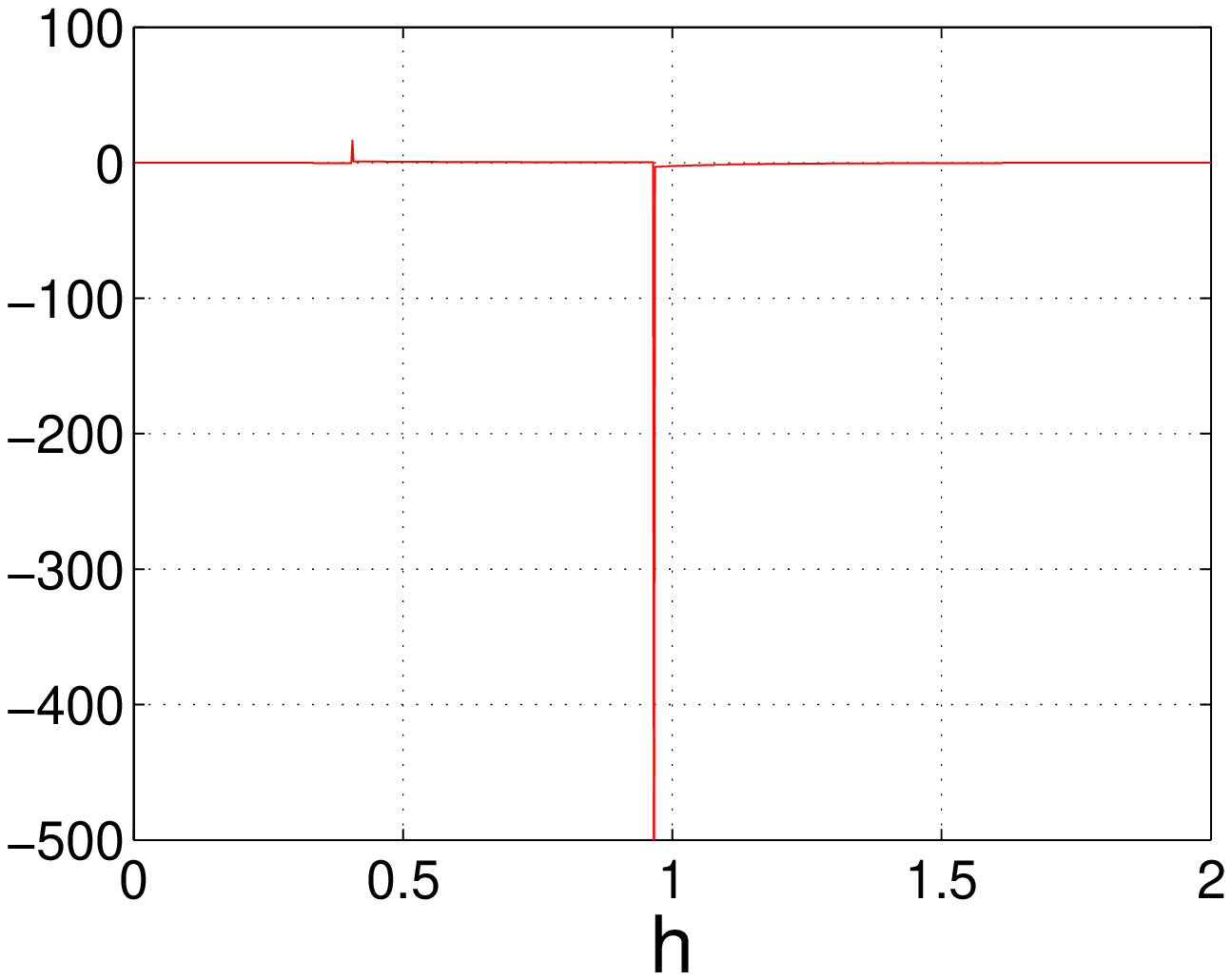}}
\subfigure{\includegraphics[width=0.225\textwidth]{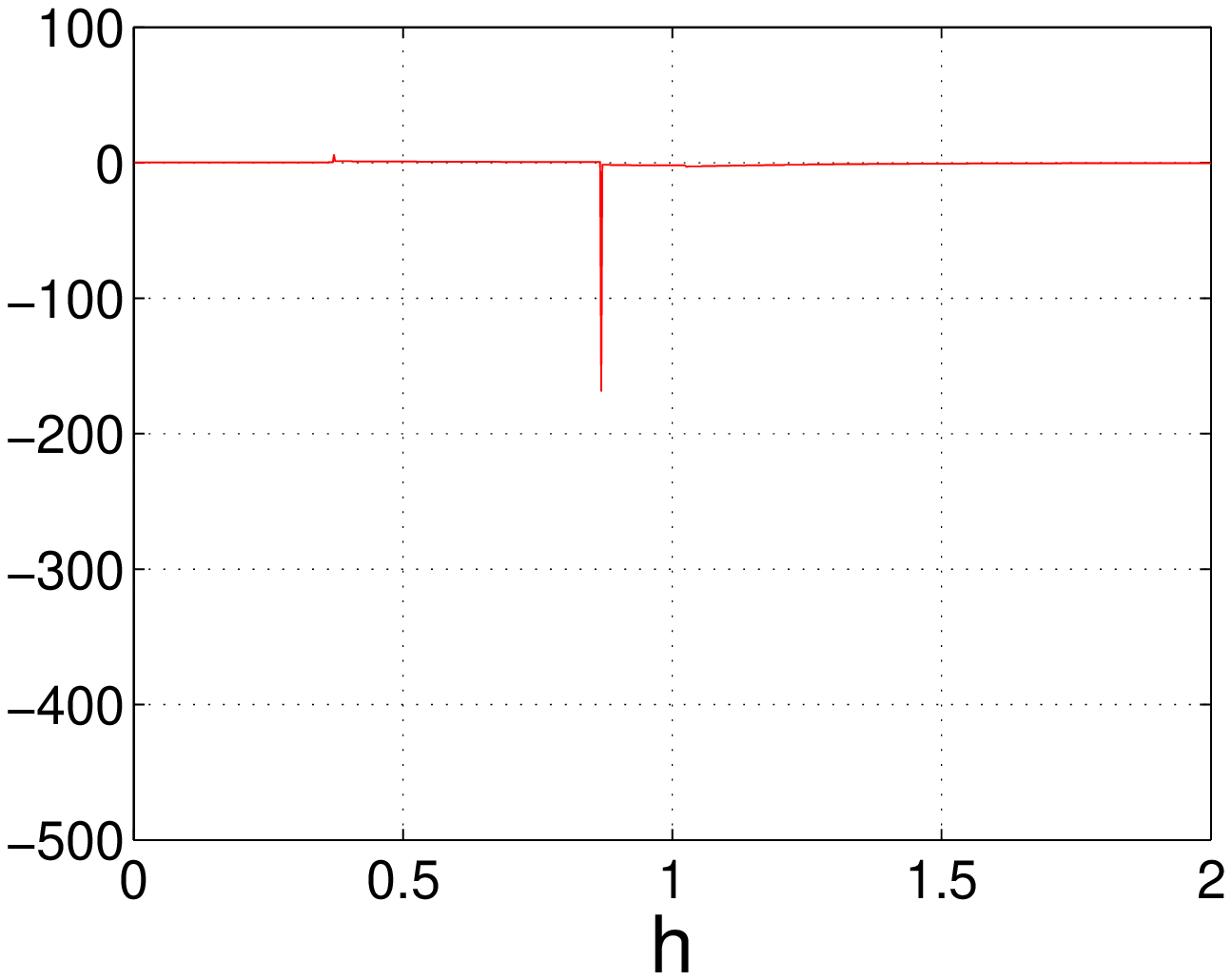}}
\subfigure{\includegraphics[width=0.225\textwidth]{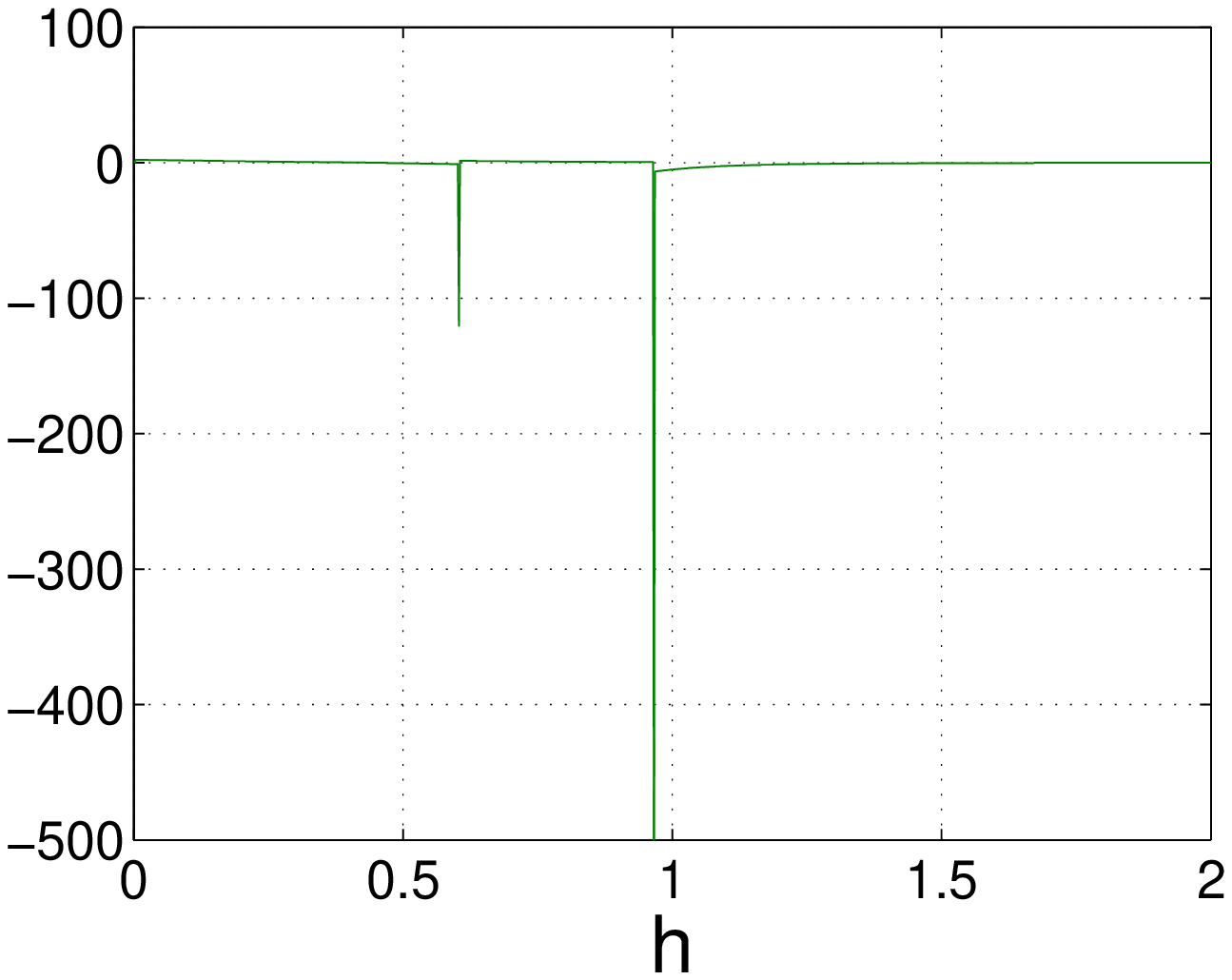}}
\subfigure{\includegraphics[width=0.225\textwidth]{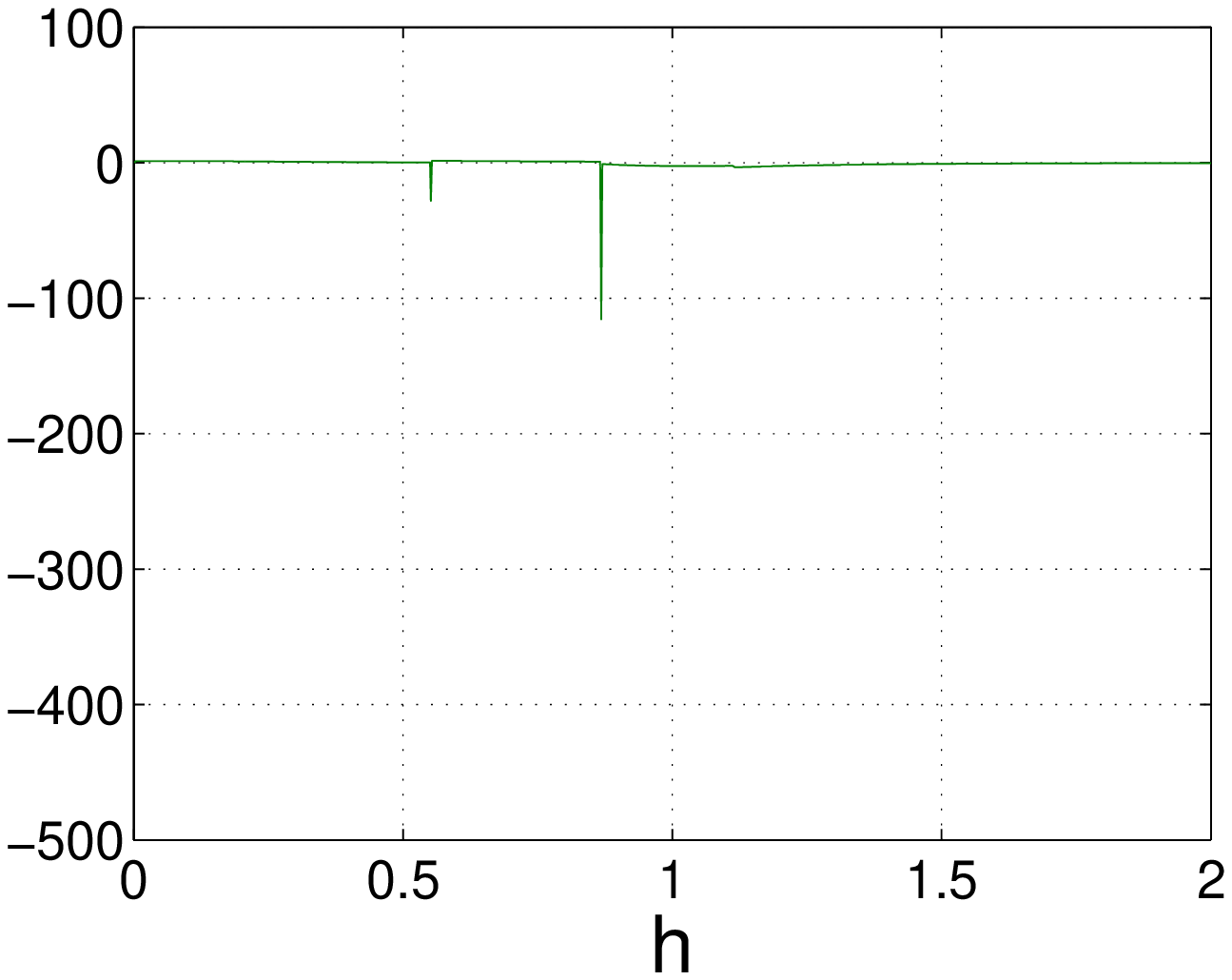}}
\caption{Derivative of the sum of all nearest neighbor bipartite GQDs.
The first column is the case that $\theta=15^{\circ}$ and the second column is for $\theta=30^{\circ}$.
 In each column from top to bottom, we investigate the system size form 3 to 5.
}
\end{figure}

In order to see the sudden change more clearly, we show the figures
of the derivative of the sum of all nearest neighbor bipartite
GQDs in Fig.4. We give 6 figures about the cases that $\theta$ equals to 15 degree and 30 degree.
The first column is the case that $\theta=15^{\circ}$ and the second column is for $\theta=30^{\circ}$.
For each case, we give three figures of $L=3,4,5$. The first
line is about $L=3$, the second line is about $L=4$, the last line
is corresponding to $L=5$. It shows that when the sudden changes
of the sum of all nearest neighbor bipartite GQDs occur, the
corresponding derivative exhibited bizarre behavior. Based on our
discussion above, only the rightmost peak of the derivative characterizes
the phase transition between phase 1A and phase 1B, other peaks reflect
the level-crossings that redefine the ground state of the system.
On the contrary, when we observed a phase transition like this, there
must be a peak of the derivative occurs. This result can be used to
predicting the sudden change of the corresponding correlation. It
is obvious that for the same angle, the peak is more obvious for small
$L$. In general, when we consider the different angles, the peak
is more obvious for small $\theta$.

\section{conclusions and discussion }
In this paper, we analyzed the global quantum discord and quantum
phase transitions in quantum critical systems. We developed a proposal
to describe the quantum phase transitions in quantum critical systems
by examining the behavior of global quantum discord. We applied this
proposal to the study of XY model. For the Ising phase transition
between phase 1 and phase 2, we find that the sum of all nearest
neighbor bipartite GQDs is effective and accurate for signaling the
phase transition point since it always reaches a maximum just around
the critical point. From another perspective, since GQD can be
considered as a resource for quantum information processing, our result
tells us that we can get most resource for quantum information and
computation tasks when the second-order quantum phase transition occurs.
In other words, in order to obtain more physical resource for quantum
information tasks from this system, we just need to adjust the corresponding
parameters to appropriate values. On the other hand, for the phase
transition between phase 1A and phase 1B, it shows that the sudden
change of GQD is very suitable for detecting the critical point of
this first-order phase transition. On the contrary, when we observed
a phase transition like this, there must be a sudden change occurs.
That is to say, we can get a sudden change of these global quantum
correlations just by creating a first-order phase transition in this
model. In detail, only the rightmost sudden change characterizes the
phase transition between phase 1A and phase 1B, other sudden changes
reflect the level-crossings that redefine the ground state of the
system. It shows that the sudden change will be more apparent when
we consider the smaller angle $\theta$ or smaller system size $L$.
That is to say, this method can be used to detecting the quantum phase
transition more easier and effective in these cases. When $\theta$
is small enough, both phase transitions can be characterized by the
rightmost sudden change of GQD. In order to see the sudden change
more clearly, we provide some figures of the derivative of the sum
of all nearest neighbor bipartite GQDs.

Our proposal may help further understanding both of the complicated
phenomena in quantum critical systems and the behavior of global quantum
correlations. This paper would initiate extensive studies of quantum
phase transitions from the perspective of quantum information processing.
On the other hand, it also shows that the nature of quantum phase
transitions can be applied to predicting the behavior of corresponding
quantum correlations. Since GQD is still meaningful for mixed states, this simple
but effective method is able to study finite-temperature phase transitions,
which is superior to other measure of quantum correlations.
Our method is also worth applying to other quantum critical systems.

\begin{acknowledgments}
We thank Yu Zeng for helpful discussions. This work is supported
by ``973'' program (2010CB922904), NSFC (11075126, 11031005,
11175248) and NWU graduate student innovation funded YZZ12083.
\end{acknowledgments}

\end{document}